\documentclass{elsarxiv}
\usepackage{natbib}
\usepackage{amsfonts,amssymb,amsmath,amsbsy,amscd}
\usepackage{latexsym,euscript,exscale,epic,eepic,epsfig}
\usepackage{subfigure}
\usepackage{times}[8pt]
\usepackage{color}
\newcommand{\jump}[1]{\left[\!\left[{#1}\right]\!\right]}      
\newcommand{\bigO}{{\rm O}}              
\newcommand{\cR}{c_{\text{R}}}           
\newcommand{\cS}{c_{\text{S}}}           
\newcommand{\cL}{c_{\text{L}}}           
\newcommand{\cSL}{c_{\text{S,L}}}        
\newcommand{\vS}{v_{\text{S}}}           
\newcommand{\vL}{v_{\text{L}}}           
\newcommand{\vSL}{v_{\text{S,L}}}        
\newcommand{\vi}{v_{\text{i}}}           
\newcommand{\vf}{v_{\text{f}}}           
\newcommand{\vif}{v_{\text{i,f}}}        
\newcommand{\vbar}{\overline{v}}         
\newcommand{\ai}{a_{\text{i}}}           
\newcommand{\af}{a_{\text{f}}}           
\newcommand{\ui}{u_{\text{i}}}           
\newcommand{\uf}{u_{\text{f}}}           
\newcommand{\aif}{a_{\text{i,f}}}        
\newcommand{\abar}{\overline{a}}         
\newcommand{\tS}{t_{\text{S}}}           
\newcommand{\tL}{t_{\text{L}}}           
\newcommand{\tSL}{t_{\text{S,L}}}        
\newcommand{\bS}{\beta_{\text{S}}}       
\newcommand{\bL}{\beta_{\text{L}}}       
\newcommand{\bSL}{\beta_{\text{S,L}}}    
\newcommand{\FD}{F_{\text{D}}}           
\newcommand{\dd}{\mathrm{d}}             
\newcommand{\ii}{\mathrm{i}}             
\newcommand{\ee}{\mathrm{e}}             
\newcommand{\Pf}{\mathop{\mathrm{Pf}}}   
\newcommand{\pv}{\mathop{\mathrm{p.v.}}}   
\newcommand{\defi}{\stackrel{\text{\tiny def}}{=}}
\newcommand{\sign}{\mathop{\text{sign}}} 
\newcommand{\siga}{\sigma_{\text{a}}}    
\newcommand{\sigD}{\sigma_{\text{D}}}    
\newcommand{\Ks}{K^{\text{s}}}           
\newcommand{\Ke}{K^{\text{e}}}           
\newcommand{\wm}{\widetilde{m}}          %
\newcommand{\wwm}{\widetilde{\widetilde{m}}}          %
\newcommand{\walpha}{\widetilde{\alpha}} %
\newcommand{\Wk}{W_{\rm k}}              
\newcommand{\Ws}{W_{\rm s}}              
\newcommand{\HY}{\mathbf{H}\!Y}          
\renewcommand{\Im}{\mathop{\mathrm{Im}}} 
\renewcommand{\Re}{\mathop{\mathrm{Re}}} 
\begin{document}
\begin{frontmatter}
\begin{minipage}{180mm}
\title{
Screw and edge dislocations with time-dependent core width:\\ From dynamical core equations to an equation of motion
}
\author[dif]{Yves-Patrick Pellegrini}
\ead{yves-patrick.pellegrini@cea.fr}
\address[dif]{CEA, DAM, DIF, F-91272 Arpajon, France.}
\end{minipage}


\begin{abstract}
\begin{minipage}{180mm}
Building on ideas introduced by Eshelby in 1953, and on recent dynamical extensions of the Peierls model for screw and edge dislocations, an approximate equation of motion (EoM) to govern non-uniform dislocation motion under time-varying stress is derived, allowing for time variations of the core width. Non-local in time, it accounts for radiative visco-inertial effects and non-radiative drag. It is completely determined by energy functions \emph{computed at constant velocity}. Various limits are examined, including that of vanishing core width. Known results are retrieved as particular cases. Notably, the EoM reduces to Rosakis's \emph{Model I} for steady motion [Rosakis, P., 2001. Supersonic dislocation kinetics from an augmented Peierls model. Phys.\ Rev.\ Lett.\ 86, 95--98]. The frequency-dependent effective response coefficients are obtained within the linearized theory, and the dynamical self-force is studied for abrupt or smooth velocity changes accompanied by core variations in the full theory. A quantitative distinction is made between low- and high-acceleration regimes, in relation to occurrence of time-logarithmic behavior.
\end{minipage}
\end{abstract}
\end{frontmatter}
\section{Introduction}
\label{sec:Introduction}
Plastic deformation in crystals arises as dislocations move through the material under an applied stress (e.g., Hirth and Lothe, 1982). Their individual motion is expected to be determined by an equation of motion (EoM), one non-trivial part of which is the \emph{self-force} (Eshelby, 1953; Ni and Markenscoff, 2008). The latter can be seen as the counterpart for defects in crystals of the Newtonian inertial force for particles. Dislocation inertia is partly grasped starting from the fact that the field configuration of the material displacement induced by a stationary dislocation is velocity-dependent (Frank, 1949; Eshelby, 1949). As the velocity $v$ changes, fields are updated to comply with the new state of motion. This updating proceeds at finite wave speed via sound wave emission from the dislocation core, which induces inertia. Updating continuously occurs over time in accelerated or decelerated motion, and because a non-supersonic dislocation always moves in its own updating wave field, dislocation inertia is history-dependent, i.e., non-local in time (Eshelby, 1953; Nabarro, 1967; Hirth and Lothe, 1982, p.\ 195). Its long-time remanent character, for \emph{infinite} rectilinear dislocations, is due to peculiarities of the associated two-dimensional wave propagation problem ---the ``afterglow'' effect discussed by Barton (1989) and Lazar (2011).

The specific question of the self-force and the related question of radiative drag involves dynamical fields generated by dislocations. These and related acoustic emission phenomena (Sedgwick, 1968) have been considered by a number of authors, notably in connection to strong-motion studies in seismology, e.g., Lund (1986) and references therein. To cite but a few analytical works, expressions of displacement, strain, or stress fields generated by arbitrarily moving screw or edge dislocations were given for an isotropic medium by Kiusalaas and Mura (1964a; 1964b), and recently revisited by Lazar (2011). Brock (1983, 1986) considered non-uniform motion along non-planar paths, in connection with crack extension. Dynamical fields produced by a finite dislocation segment have been addressed by Lund (1986). Arias and Lund (1999) studied motion in samples of finite width. Beltzer (1982) considered dynamical acoustic emission associated to random dislocation motion. The self-force itself has been explored after Eshelby by several authors, notably Beltz et al. (1968), and more recently by Markenscoff and co-workers who undertook a systematic study of the singularities associated to moving dislocations (Markenscoff, 1980; Ni and Markenscoff, 2008; Markenscoff and Huang, 2008).

Consideration of systems with lattice periodicity leads to involved dispersion relations (e.g., Askar, 1986), with incidence on dislocation dynamics (Eshelby, 1956). For alternative approaches using gradient elasticity see, e.g., Eringen (2002) and references therein, and Lazar (2010). However, as far as inner length scales are concerned, lattice-related features can approximately be accounted for in classical continuum elasticity via a Peierls--Nabarro (PN) cohesive-zone approach on a prescribed slip plane, which produces dislocations of finite width without introducing additional wavemodes. Accordingly, we restrict ourselves to planar cores, i.e., Somigliana dislocations (Nabarro, 1967), such as in FCC crystals (Heidenreich and Shockley, 1948).

In this context, this paper addresses the question of obtaining the EoM, with focus on its \emph{structure}, thus trying to answer a question implicitly raised by the phenomenological approach of Pillon et al.\ (2007). The latter suggests that an EoM of simple form, valid for all velocity regimes under minimal assumptions, might be within reach. The present work, to be seen as an effort to notably reduce the part of phenomenology of the latter work, answers positively. To this aim, we start from dynamical extensions of the well-known static PN equation (Peierls, 1940; Nabarro, 1947; Schoeck 2005) for the shape of dislocation cores. Following initial steps taken by Eshelby (1953) and, e.g.,  Stenzel (1969), such \emph{dynamical core equations} were recently obtained for screw and edge dislocations  (Pellegrini, 2010; 2011) by means of the Green's function approach (Mura, 1987) in the framework of generalized-function theory. No attempt will be made to solve numerically the EoM to be obtained under an applied stress, which is left to future work. We instead try to gain some fresh insight into its most important component, the self-force, by artificially prescribing the motion, which is usual practice in this kind of studies. The embedding medium is infinite, although finite-size effects may notably influence fast dislocation motion (Arias and Lund, 1999; Vandersall and Wirth, 2004).

The dynamical core equations are reviewed in Section \ref{sec:DynPN}. The method used to derive from them the EoM is presented in Section \ref{sec:eomprinc}. Arbitrary time-variation of core width is permitted, a notable difference with Eshelby's treatment (1953). In Section \ref{sec:selfforce} the EoM is made fully explicit by means of identities further discussed in Appendix A. It involves only known \emph{stationary} energy functions, which simplifies matters in formally bringing inertial effects for edges and screws down to the same level of complexity. The linearized theory is considered in Section \ref{sec:fdem}. Section \ref{sec:counterterm} provides further insight in the Volterra limit of vanishing core width, in connection with jumps between two arbitrary velocities. Core-width velocity dependence is discussed in Section \ref{sec:tdcw}. In turn, Section \ref{sec:generalization} addresses velocity changes, either abrupt or smooth in association with variations of core width, and analyzes the possible regimes of the self-force. Section \ref{sec:asym} specializes the discussion to the logarithmic regime. We summarize our findings and conclude in Section \ref{sec:concl}. Appendix B gathers some useful integrals. Details on numerical calculations are provided in Appendix C.

Our conventions are as follows. Shear and longitudinal sound velocities are denoted by $\cS$ and $\cL$. A screw dislocation is either subsonic ($|v|<\cS$) or supersonic ($|v|>\cS$), whereas an edge dislocation is either subsonic ($|v|<\cS$), transonic ($\cS<|v|<\cL$) or supersonic ($|v|>\cL$). Time and space Fourier transforms (FT) are carried out as
\begin{eqnarray}
f(x,t)=\int\frac{\dd k}{2\pi}\frac{\dd \omega}{2\pi}\ee^{\ii(kx-\omega t)}f(k,\omega),
\end{eqnarray}
$f(x,t)$ being some function of position $x$ and time $t$. Following usual practice, we distinguish it from its FT $f(k,\omega)$ only by the symbols of the Fourier momentum $k$ and the angular frequency $\omega$. Integrals with implicit bounds run from $-\infty$ to $+\infty$.
\section{Dynamical core equations}
\label{sec:DynPN}
\subsection{Overview}
\label{sec:overview}
Let the rectilinear dislocation glide in the plane $y=0$ of an isotropic medium of infinite extent. The abscissa in the slip direction is $x$. Let furthermore $\eta(x,t)$ be the relative material displacement (slip) between both sides of the slip plane. Given the function  $\siga(x,t)$ that represents a prescribed resolved shear stress applied on the slip plane by external agents, the governing \emph{dynamical core equation} for $\eta(x,t)$ is the stress balance equation
\begin{subequations}
\begin{eqnarray}
\label{eq:pndyn}
\sigma_\eta(x,t)+\sigD(x,t)+\siga(x,t)=f'\left(\eta(x,t)\right),
\end{eqnarray}
where $-f'(\eta)$ is the lattice pull-back force that derives from the lattice potential  $f(\eta)$, usually identified to the $\gamma$-potential (Christian and Vitek, 1970). This function is $b$-periodic, with $b$ the Burgers vector length along the slip direction, and such that $f'(0)=0$. The function $\sigma_\eta(x,t)$ represents the dynamical \emph{self-stress} produced by the moving dislocation on the slip plane, and reads (Pellegrini, 2010; 2011)
\begin{eqnarray}
\label{eq:pnstress}
\sigma_\eta(x,t)\defi-\frac{\mu}{\pi}\int\dd \tau\,\dd x'\,K(x,t|x',\tau)\frac{\partial
\eta}{\partial x}(x',\tau)-\frac{\mu}{2\cS}\frac{\partial\eta}{\partial t}(x,t),
\end{eqnarray}
where $\mu$ is the shear modulus and $K$ is a non-local kernel that depends on the dislocation character. Hereafter, the term proportional to $\partial\eta/\partial t$, local in space and time, is referred to as the \emph{local term}. Finally, the stress
\begin{eqnarray}
\label{eq:dragstress}
\sigD(x,t)\defi -\alpha\frac{\mu}{\cS}\frac{\partial\eta}{\partial t}(x,t)
\end{eqnarray}
\end{subequations}
has the same form as the local term in $\sigma_\eta$. It represents a phenomenological drag component, with positive dimensionless damping coefficient $\alpha$, which is intended to account for non-radiative damping effects (Movchan et al., 1998) and proves necessary for compatibility with a non-zero applied stress in the steady state (Rosakis, 2001). We focus on single-dislocation solutions such that $\eta(-\infty,t)=b+\eta(+\infty,t)$ with
$\eta(+\infty,t)$ such that $\siga(+\infty,t)=f'\left(\eta(+\infty,t)\right)$, assuming for consistency that $\siga(+\infty,t)=\siga(-\infty,t)$. We refer to Pellegrini (2010) for the specification of which component of the stress tensor the quantity $\sigma_\eta$ corresponds to, which depends on the dislocation character.

Expressions for $K$ are as follows. For a screw dislocation,
\begin{eqnarray}
\label{eq:kxt}
K(x,t|x',\tau)=\Ks(x-x',t-\tau)\quad\text{with}\quad \Ks(x,t)\defi\frac{x}{2\cS t^2}\frac{\theta(\cS t-|x|)}{\sqrt{\cS^2
t^2-x^2}},
\end{eqnarray}
where $\theta(x)$ is Heaviside's unit-step function. Instead, for a gliding edge dislocation
\begin{subequations}
\label{eq:noyauxcg}
\begin{eqnarray}
\label{eq:kxte}
K(x,t|x',\tau)&=&\Ke_1(x-x',t-\tau)+\Ke_2(x-x',t-\tau)\frac{\partial}{\partial x'},
\end{eqnarray}
with
\begin{eqnarray}
\label{eq:noyauxcg1}
\Ke_1(x,t)&\defi&\frac{2\cS^2}{x^3}
\left[\frac{1}{\cL}\frac{2\cL^2
t^2-x^2}{\sqrt{\cL^2 t^2-x^2}}\theta(\cL t-|x|)-\frac{1}{\cS}\frac{2\cS^2 t^2-x^2}{\sqrt{\cS^2 t^2-x^2}}\theta(\cS t-|x|)\right]
+\frac{x}{2\cS t^2}\frac{\theta(\cS t-|x|)}{\sqrt{\cS^2 t^2-x^2}},\\
\label{eq:noyauxcg2}
\Ke_2(x,t)&\defi&\frac{\cS}{2}\frac{\theta(\cS t-|x|)}{\sqrt{\cS^2 t^2-x^2}}.
\end{eqnarray}
\end{subequations}
These causal kernels were derived using a retarded Green function of the material displacement and vanish for $t<0$. As functions of $x$, they are regular at $x=0$ for $t\not=0$. Because $\Ke_1(x,t)$ is an $\bigO(x)$, it is preferable in the perspective of integrating over $x$ to emphasize this regularity by writing it in the following form where all terms are separately well-defined at $x=0$ if $t\not=0$:
\begin{eqnarray}
\label{eq:noyauxcg1bis}
\Ke_1(x,t)&=&\frac{2\cS^2}{x^3}
\left[\frac{1}{\cL}\left(\frac{2\cL^2
t^2-x^2}{\sqrt{\cL^2 t^2-x^2}}-2\cL t\right)\theta(\cL t-|x|)
-\frac{1}{\cS}\left(\frac{2\cS^2 t^2-x^2}{\sqrt{\cS^2 t^2-x^2}}-2\cS t\right)\theta(\cS t-|x|)\right]\nonumber\\
&&{}+4\frac{\cS^2 t}{x^3}\left[\theta(\cL t-|x|)-\theta(\cS t-|x|)\right]+\frac{x}{2\cS t^2}\frac{\theta(\cS t-|x|)}{\sqrt{\cS^2 t^2-x^2}}.
\end{eqnarray}
With expressions (\ref{eq:kxt}) and (\ref{eq:noyauxcg}) of the kernels, Eq.\ (\ref{eq:pndyn}) has been shown (Pellegrini, 2010) to reduce, in the steady-state regime, to Weertman's equations\footnote{Namely, velocity-dependent stationary generalizations of the PN model.} (1969) in an augmented form with additional drag term. The latter extension of the PN model has been introduced by Rosakis (2001) under the name \emph{Model I}.

In the above equations, $\theta(1-|x|)/\sqrt{1-x^2}$ is merely a particular notation for the composition of $1-x^2$ with the locally-integrable function (distribution) $x_+^{-1/2}$ equal to $x^{-1/2}$ if $x>0$ and $0$ if $x<0$. Consider more generally the locally-integrable function $x_+^{-\alpha}\defi\{x^{-\alpha}$ if $x>0$; $0$ if $x<0\}$ where $\Re\alpha<1$. Its derivative is the pseudo-function $
\left(x_+^{-\alpha}\right)'=-\alpha\Pf x_+^{-\alpha-1}$,
where $\Pf$ stands for Hadamard's finite part (Schwartz, 1966, p.\ 38). With the shorthand notations
\begin{eqnarray}
\label{eq:vsldef}
\vSL\defi x/(\cSL t),
\end{eqnarray}
the kernel $\Ks$ in Eq.\ (\ref{eq:kxt}) takes the self-similar form
\begin{subequations}
\begin{eqnarray}
\label{eq:ksalt}
\Ks(x,t)=\frac{\theta(t)}{2\cS t^2}\vS (1-\vS^2)_+^{-\frac{1}{2}},
\end{eqnarray}
where a Heaviside factor $\theta(t)$ is introduced to preserve causality in this alternative writing. Moreover, since in the sense of distributions
\begin{eqnarray*}
\Ke_2(x-x',t-\tau)\frac{\partial}{\partial x'}=-\frac{\partial \Ke_2}{\partial x'}(x-x',t-\tau)=\frac{\partial \Ke_2}{\partial x}(x-x',t-\tau),
\end{eqnarray*}
by use of $\left(x_+^{-\alpha}\right)'=-\alpha\Pf x_+^{-\alpha-1}$ with $\alpha=1/2$, the kernel $K$ in Eq.\ (\ref{eq:kxte}) is alternatively written as
$K(x,t|x',\tau)=\Ke(x-x',t-\tau)$ with
\begin{eqnarray}
\label{eq:kealt}
\Ke(x,t)&\defi&\frac{\theta(t)}{2\cS t^2}\left\{\frac{4}{\vS^3}
\left[(2
-\vL^2)(1-\vL^2)_+^{-\frac{1}{2}}-(2-\vS^2)(1-\vS^2)_+^{-\frac{1}{2}}
\right]
+\vS(2-\vS^2)\Pf (1-\vS^2)_+^{-\frac{3}{2}}\right\}.
\end{eqnarray}
\end{subequations}
Although we shall proceed otherwise in Section \ref{sec:eomprinc} (namely, by using the kernel in its form (\ref{eq:kxte})), introducing a finite part prescription at this step is quite natural, and makes $\Ke$ well-defined at $\vS=1$.

Still, due to their self-similarity in $x$ and $t$, the kernels are ambiguous at $(x,t)=(0,0)$ where the ratio $x/t$ is not defined. The way to go to this limit is a crucial issue in the case of a Volterra dislocation, which has no intrinsic length scale. The writings of Eqs.\ (\ref{eq:ksalt}b) put emphasis on an interpretation of the kernels as functions of the ``velocities'' $\vSL$ defined in (\ref{eq:vsldef}), and of time. However, this interpretation is possible only if $t\not=0$. As a consequence the kernels must be regularized at $t=0$. This is conveniently achieved by multiplying them by a factor $\ee^{-\epsilon/t}$, where $\epsilon$ is a positive constant that sets an ``inner'' time scale, and by taking the limit $\epsilon\to 0$ after the time integral over $\tau$ is done in Eq.\  (\ref{eq:pnstress}). Unless otherwise mentioned, this prescription is implicit henceforth; see Pellegrini (2011) for additional information.

\subsection{The local term in $\sigma_\eta$}
\label{sec:lossterm}
The complementarity of the regularized non-local kernel $K$ and of the local term in Eq.\ (\ref{eq:pnstress}) is illustrated in a simple non-stationary case where known results are retrieved. Consider a Volterra screw dislocation located at position $\xi(t)$, at rest with $\xi(t)=0$ for $t<0$ and moving arbitrarily at times $t>0$. For such a dislocation,
\begin{eqnarray}
\eta(x,t)=b\theta(\xi(t)-x).
\end{eqnarray}
Using this expression in (\ref{eq:pnstress}) with kernel (\ref{eq:kxt}), the stress $\sigma_\eta$ becomes such that (Pellegrini, 2010)
\begin{eqnarray}
\label{eq:stressy0}
\frac{2\pi\sigma_\eta}{\mu b}&=&
\frac{1}{c_{\text{S}}}\int_0^t\frac{\text{d}\tau}{(t-\tau)^2}\frac{\vbar}{\cS}
\left[1-\left(\frac{\vbar}{\cS}\right)^2\right]_+^{-\frac{1}{2}}-\frac{\pi}{c_{\text{S}}}\delta(\xi-x)\dot{\xi}
+\left\{1-\left[1-\left(\frac{x}{\cS t}\right)^2\right]_+^{\frac{1}{2}}\right\}\frac{1}{x}\quad (t>0),
\end{eqnarray}
where
\begin{eqnarray}
\vbar(x,t,\tau)\defi\frac{x-\xi(\tau)}{t-\tau}.
\end{eqnarray}
The Dirac contribution comes from the local term, and the last term is the integrated contribution of times $\tau<0$ prior to motion.

Now, represent the function $x(1-x^2)_+^{-\frac{1}{2}}$ by its Fourier transform $-\ii\pi J_1(k)$, where $J_1$ is the Bessel function. With the regularization procedure described above, replacing for convenience $\epsilon$ by $\epsilon/\cS$, one gets
\begin{eqnarray}
&&\int_0^t\text{d}\tau\frac{ \ee^{-\frac{\varepsilon}{\cS(t-\tau)}}}{\cS(t-\tau)^2}\frac{\vbar}{\cS}
\left[1-\left(\frac{\vbar}{\cS}\right)^2\right]_+^{-\frac{1}{2}}
=-\ii\pi\int\frac{\text{d}k}{2\pi}\, J_1(k)\int_0^t\text{d}\tau\frac{\ee^{\ii\frac{k[x-\xi(\tau)]+i\epsilon}{\cS(t-\tau)} }}{\cS(t-\tau)^2}\nonumber\\
\label{eq:res1}
\end{eqnarray}
To make progress, we specialize to the case where the dislocation jumps from rest to a steady state of velocity $v$ for $t>0$. Then with $\xi(t)=vt$,
\begin{eqnarray}
\int_0^t\text{d}\tau\frac{\ee^{\ii\frac{k[x-\xi(\tau)]+i\epsilon}{\cS(t-\tau)} }}{\cS(t-\tau)^2}&=&\ee^{\ii k \frac{v}{\cS}}\int_0^t\text{d}\tau\,\frac{\ee^{\ii \frac{k(x-vt)+i\epsilon}{\cS(t-\tau)}}}{\cS(t-\tau)^2}.
\end{eqnarray}
For $x$, $k$ $\in\mathbb{R}$, using the change of variables $u=1/[\cS(t-\tau)]$ leads to
\begin{eqnarray}
&&\lim_{\epsilon\to 0^+}\int_0^t\text{d}\tau\,\frac{\ee^{\ii \frac{k x+i\epsilon}{\cS(t-\tau)}}}{\cS(t-\tau)^2}
=\lim_{\epsilon\to 0^+}\ii\frac{\ee^{\ii \frac{k x}{\cS t}}}{kx +\ii\epsilon}=\ii\,\ee^{\ii \frac{k x}{\cS t}}\mathop{\text{p.v.}}\frac{1}{kx}+\pi\delta(kx),
\end{eqnarray}
where $\text{p.v.}$ is the principal value. Since $J_1(k)/k$ is finite at $k=0$ and $\delta(a x)=\delta(x)/|a|$ for $a\in\mathbb{R}$, $a\not=0$, one has
\begin{eqnarray}
\label{eq:res2}
\lim_{\epsilon\to 0^+}J_1(k)\int_0^t\text{d}\tau\,\frac{\ee^{\ii \frac{k x+i\epsilon}{\cS(t-\tau)}}}{\cS(t-\tau)^2}
=\ii\frac{J_1(k)}{k}\ee^{\ii \frac{k x}{\cS t}}\left(\mathop{\text{p.v.}}\frac{1}{x}\right)+\pi\frac{J_1(k)}{|k|}\delta(x)
\end{eqnarray}
so that the limiting value $k=0$ causes no trouble. Applying this result to Eq.\ (\ref{eq:res1}), and using the fact that $J_1(k)/k$ and $J_1(k)/|k|$ are even and odd, respectively, yields
\begin{eqnarray}
&&\lim_{\epsilon\to 0^+}\int_0^t\text{d}\tau\frac{\ee^{-\frac{\varepsilon}{\cS(t-\tau)}}}{\cS(t-\tau)^2}
\frac{\vbar}{\cS}
\left[1-\left(\frac{\vbar}{\cS}\right)^2\right]_+^{-\frac{1}{2}}\nonumber\\
\label{eq:res3}
&=&-\ii\pi\left\{
\ii\left(\mathop{\text{p.v.}}\frac{1}{x-vt}\right)\int_0^\infty \frac{\dd k}{\pi}\frac{J_1(k)}{k}\cos\left(\frac{kx}{\cS t}\right)
+\ii\pi\delta(x-vt)\int_0^\infty \frac{\dd k}{\pi}\frac{J_1(k)}{k}
\sin\left(\frac{kv}{\cS}\right)\right\}.
\end{eqnarray}
The last step consists in making Eq.\ (\ref{eq:res3}) explicit by invoking the following integrals (Gradshteyn and Ryzhik, 2007):
\begin{subequations}
\begin{eqnarray}
\label{eq:cosj}
\int_0^\infty \frac{\dd k}{\pi}\frac{J_1(k)}{k}\cos(k x)&=&(1-x^2)_+^{\frac{1}{2}},\\
\label{eq:sinj}
\int_0^\infty \frac{\dd k}{\pi}\frac{J_1(k)}{k}\sin(k x)&=&x-(x^2-1)_+^{\frac{1}{2}}\sign(x).
\end{eqnarray}
\end{subequations}
Using the outcome in Eq.\ (\ref{eq:stressy0}) and reorganizing terms, the Dirac contribution of the local term eventually cancels out thanks to the linear term, $x$, in Eq.\ (\ref{eq:sinj}). The self-stress on the slip plane follows as
\begin{eqnarray}
\label{eq:stressy2}
\frac{2\pi}{\mu b}\sigma_\eta(x,t)&=&\frac{1}{x}\left\{1+\mathop{\text{p.v.}}\frac{vt}{x-vt}
\left[1-\left(\frac{x}{\cS t}\right)^2\right]_+^{\frac{1}{2}}
\right\}
-\pi\sign(v)\left[\left(v/\cS\right)^2-1\right]_+^{\frac{1}{2}}\,\delta(x-vt).
\end{eqnarray}
Apart from an overall minus sign due to a different choice of dislocation sign,
the right-hand side (rhs) of Eq.\ (\ref{eq:stressy2}) reduces when $|v|<\cS$ to the stress generated by a subsonic Volterra dislocation. Except for the principal value prescription (Pellegrini, 2011), an equivalent expression is found (up to a factor $\mu$) in Eq.\ (17) of Markenscoff (1980). For $|v|>\cS$ we also retrieve Eq.\ (2) of Callias and Markenscoff (1980) at $z=0$ in their notations. The Dirac term at the dislocation position $x=vt$, present only for $|v|>\cS$, indicates a coincidence between this position and the tip of the Mach front generated by the dislocation in this supersonic regime (Callias and Markenscoff, 1980).

The cancelation that takes place between the local term and a contribution arising from the integral kernels in the above calculation is reexamined in Section \ref{sec:counterterm} from a different point of view, namely, at the level of the EoM where the dislocation has a finite core width $a$. As this introduces an inner time scale in the problem, working out explicitly the $\epsilon$-regularization will prove superfluous.

\section{Equation of motion: principle}
\label{sec:eomprinc}
An equation of motion for the dislocation position $\xi(t)$ is obtained by
eliminating the degrees of freedom of the core shape out of the dynamical core equation.
To this aim, one multiplies this equation by
$\rho(x,t)\defi(\partial\eta/\partial x)(x,t)$, where $\eta(x,t)$ is its solution,
and integrates over $x$, which results in a work balance equation. Because the problem is one-dimensional the latter equation is understood as the EoM for $\xi(t)$  (Eshelby, 1953). Although $\rho$ is unknown, it is expected that in a first approximation the EoM mainly depends on $\xi(t)$ and on the core width, $a(t)$, rather than of the exact shape of $\rho$. As a substitute to the solution $\eta$ of the core equation, we follow Eshelby and appeal to the usual arctangent \textit{ansatz}, which solves the steady-state equation for a sine pull-back force law (Weertman, 1969), but only for subsonic velocities (Eshelby, 1956; Weertman, 1967). Accordingly, what follows is not meant to apply to supersonic velocities \textit{stricto sensu} although we shall occasionally refer to the latter to the purpose of enlightening some aspects of the equations.

The Lorentzian $\rho(x,t)/b$ can be considered as a delta-sequence to approach the Volterra limit of zero core width (Ni and Markenscoff, 2008). Here, however, the value of $a$ is of physical interest. For that reason, we relax Eshelby's original rigid core assumption (Eshelby, 1953) by allowing for a time-dependent core width $a(t)$, the latter dependence being left arbitrary for the time being. We thus take
\begin{eqnarray}
\label{eq:etaansatz}
\eta(x,t)=\eta_0(t)+\frac{b}{\pi}\left[\frac{\pi}{2}
-\arctan\frac{2\left(x-\xi(t)\right)}{a(t)}\right],
\end{eqnarray}
where $\xi(t)$ is the position of the dislocation center, and
$\eta(+\infty,t)=\eta_0(t)$. We use henceforth the shorthand notation
$v(t)\defi \dot{\xi}(t)$ for the instantaneous velocity. The domain of relevance of this \textit{ansatz} is not clearly established. Numerical comparisons with phase-field calculations (Pillon et al., 2007) have indirectly shown that the arctangent approximates well the true time-dependent solution in the subsonic range, for velocities less than the Rayleigh velocity $\cR$ for edges. No such confirmation is available for $\cR<|v|<\cL$ and ---independently of the core shape issue--- problems that arise for $|v|>\cR$ are evoked in Section \ref{sec:tdcw}, in relation to the model we shall adopt for $a(t)$.

Assume that the spatial scale of variation of the driving stress $\siga(x,t)$
is much larger than the core size. The applied stress averaged over the core, $\siga(t)$, is approximately
\begin{eqnarray}
\label{eq:pkforce}
\siga(t)\defi-\frac{1}{b}\int \dd x\,\rho(x,t)\siga(x,t)
\simeq \siga\bigl(\xi(t),t\bigr),
\end{eqnarray}
and $b\siga(t)$  is the Peach-Koehler force.
On the other hand, the self-force of the dislocation is
\begin{eqnarray}
\label{eq:fesh_gen}
F(t)&\defi&\int \dd x\,\rho(x,t)\sigma_\eta(x,t)
=-\frac{\mu}{\pi}\int\dd \tau\,\dd x\,\dd x'\,
\rho(x,t) K(x,t|x',\tau)\rho(x',\tau)+\frac{\mu b^2}{2\pi \cS}
\frac{v(t)}{a(t)},
\end{eqnarray}
where $\sigma_\eta(x,t)$ is read in Eq.\ (\ref{eq:pnstress}) and where the local contribution,
\begin{eqnarray}
-\frac{\mu}{2\cS}\int \dd x\,\rho(x,t)\frac{\partial\eta}{\partial t}(x,t)=\frac{\mu b^2}{2\pi \cS}
\frac{v(t)}{a(t)}
\end{eqnarray}
has been evaluated using
\begin{eqnarray}
\frac{\partial\eta}{\partial t}(x,t)=-\left\{v(t)+[x-\xi(t)]\frac{\dot{a}(t)}{a(t)}\right\}\rho(x,t)
\end{eqnarray}
and the identities $\int \dd x \rho^2(x,t)=b^2/[\pi a(t)]$ and  $\int \dd x [x-\xi(t)]\rho^2(x,t)=0$. The latter three equations are consequences of Eq.\ (\ref{eq:etaansatz}). By the same token, the phenomenological ``viscous'' stress $\sigD$ in Eq.\ (\ref{eq:dragstress}) gives rise to a drag force
\begin{eqnarray}
\label{eq:dragforce}
\FD(t)=2\alpha\frac{\mu b^2}{2\pi \cS}
\frac{v(t)}{a(t)}.
\end{eqnarray}
Due to relativistic effects $a$ depends in general on the velocity, so that this drag force is nonlinear in $v$ (see Section \ref{sec:tdcw}).

Finally, because $f$ is $b$-periodic and $\eta(-\infty,t)=b+\eta(+\infty,t)$,
\begin{eqnarray}
\int \dd x\,\rho(x,t)f'(\eta(x,t))=f(\eta(+\infty,t))-f(\eta(-\infty,t))=0.
\end{eqnarray}
Thus, the core equation projected on $\rho$ becomes the force balance equation
\begin{eqnarray}
\label{eq:eomgen}
F(t)+\FD(t)=b\siga(t).
\end{eqnarray}
Since $F(t)$ is a functional of $\xi$ over its history,
this equation constitutes the desired EoM for $\xi$. The self-force is tantamount to a visco-inertial drag force (Nabarro, 1951; Eshelby, 1953). The following focuses on obtaining its expression in a physically appealing form, and on examining some consequences.

For practical purposes, it is useful to transform the double integral over $x$ and $x'$ in Eq.\ (\ref{eq:fesh_gen}) in the following way. Let $\rho(k,t)$ be the spatial FT of $\rho(x,t)$. Setting
$\Delta x\defi x-x'$ and $\Delta t\defi t-\tau>0$, one has for a translation-invariant kernel $K_1$ of the $\Ks$ or $\Ke_1$ types
\begin{eqnarray}
\int\dd x\,\dd x'\,\rho(x,t)&& K_1(x-x',t-\tau)\rho(x',\tau)
=
\int\dd \Delta x\, K_1(\Delta x,\Delta t)\int\frac{\dd k}{2\pi}\,
\rho(k,t)\rho(-k,\tau)\ee^{\ii k \Delta x}.
\end{eqnarray}
Introduce now the \emph{average complex velocity} between instants $t$ and $\tau$ defined as
\begin{eqnarray}
\label{eq:vbar}
\vbar(t,\tau)\defi\frac{\xi(t)-\xi(\tau)}{\Delta t}+\ii\frac{\abar(t,\tau)}{\Delta t},
\end{eqnarray}
where $\abar(t,\tau)\defi[a(t)+a(\tau)]/2$. With the \textit{ansatz} (\ref{eq:etaansatz}), one has
\begin{eqnarray}
\label{eq:rhok}
\rho(k,t)=-b \ee^{-\ii\xi(t)k-\frac{1}{2}a(t)|k|}.
\end{eqnarray}
Since $\Im \vbar>0$, one finds
\begin{eqnarray}
\int\frac{\dd k}{2\pi}\,\rho(k,t)\rho(-k,\tau)\ee^{\ii k \Delta x}&=&\frac{b^2}{\pi}\Re\int_0^\infty \dd k\,\ee^{-\ii(\Delta x-\Delta t\, \vbar)k}
=\frac{b^2}{\Delta t}\Re\left(\frac{1}{\ii\pi}\frac{1}{\frac{\Delta x}{\Delta t}-\vbar(t,\tau)}\right),
\end{eqnarray}
whence
\begin{eqnarray}
\label{eq:firstint}
\int\dd x\,\dd x'\,\rho(x,t)&& K_1(x-x',t-\tau)\rho(x',\tau)=b^2\Re\int\dd \frac{\Delta x}{\Delta t}
\frac{K_1(\Delta x,\Delta t)}{\ii\pi\left[\frac{\Delta x}{\Delta t}-\vbar(t,\tau)\right]}.
\nonumber\\
\end{eqnarray}
In a similar way, since $\partial/\partial x'=-\partial/\partial\Delta x$, one has
\begin{eqnarray}
\label{eq:secondint}
&&\int\dd x\,\dd x'\,\rho(x,t)\Ke_2(x-x',t-\tau)\frac{\partial\rho}{\partial x'}(x',\tau)
=\frac{b^2}{\Delta t}\Re\int\dd \frac{\Delta x}{\Delta t}
\frac{\Ke_2(\Delta x,\Delta t)}{\ii\pi\left[\frac{\Delta x}{\Delta t}-\vbar(t,\tau)\right]^2}.
\end{eqnarray}
Roughly, these integrals over $\Delta x/\Delta t$ perform the task of replacing within $K_1$ (resp., in $\partial K_2/\partial x$) the quantity $\Delta x$ by $\vbar(t,\tau)\Delta t$. We now are in position to compute $F(t)$ for arbitrarily moving dislocations.

\section{Self-force for screw and edge dislocations}
\label{sec:selfforce}
\subsection{Screw dislocation}
Invoking Eq.\ (\ref{eq:firstint}) with kernel (\ref{eq:kxt}) appropriate to
a screw dislocation, and appealing to (\ref{eq:b1}) to carry out the integration over $\Delta x/\Delta t$ by changing variables with $u=\Delta x/(\cS\Delta t)$, the generic expression (\ref{eq:fesh_gen}) of $F(t)$
takes the form
\begin{eqnarray}
F(t)=
-\frac{\mu b^2}{2\pi\cS}\Re\int_{-\infty}^t\frac{\dd \tau}{\Delta t^2}\,\frac{(\vbar/\cS)}{\sqrt{1-\vbar^2/\cS^2}}+\frac{\mu b^2}{2\pi \cS}\frac{v(t)}{a(t)},
\end{eqnarray}
where $\vbar$ stands for $\vbar(t,\tau)$. Introducing the characteristic energy per unit length of dislocation line as (e.g., Hirth et al., 1998)
\begin{eqnarray}
\label{eq:w0defmain}
w_0\defi \mu b^2/(4\pi),
\end{eqnarray}
it is realized at this point that the function
\begin{eqnarray}
\label{eq:fsvdef}
p(v)=\frac{w_0}{\cS}\frac{(v/\cS)}{\sqrt{1-v^2/\cS^2}}
\end{eqnarray}
is nothing but the quasimomentum $p=\dd L/\dd v$, where $L(v)$ is the screw Lagrangian function (see Appendix A).
Thus,
\begin{eqnarray}
\label{eq:eomexpl}
F(t)=-2\Re\int_{-\infty}^t\frac{\dd \tau}{\Delta t^2}\,p(\vbar)+\frac{\mu b^2}{2\pi \cS}\frac{v(t)}{a(t)}.
\end{eqnarray}
Introducing moreover the associated stationary mass function $m(v)$ as
\begin{eqnarray}
\label{eq:massdef}
m\defi\frac{\dd p}{\dd v},
\end{eqnarray}
one partial integration applied to the integral yields
\begin{eqnarray}
2\Re\int_{-\infty}^t\frac{\dd \tau}{\Delta t^2}\,p(\vbar)
=-2\Re\int_{-\infty}^t\frac{\dd \tau}{\Delta t}m\left(\vbar(t,\tau)\right)
\frac{\dd \vbar}{\dd \tau}(t,\tau).
\end{eqnarray}
Indeed, provided that the dislocation was stationary in the remote past, with `initial' velocity $\vi$, the boundary term $\Re[p(\vbar)/\Delta t]$ in the partial integration vanishes at $\tau=-\infty$ where $\vbar(t,\tau)=\vi+\ii 0^+$. Its vanishing at $\tau=t^-$ as well stems from the following considerations. Eq.\ (\ref{eq:vbar}) entails the expansion
\begin{eqnarray}
\label{eq:ovexpan}
\vbar(t,\tau)&=&\ii \frac{a(t)}{t-\tau}+v(t)-\frac{\ii}{2}\dot{a}(t)+\bigO(\tau-t)
\end{eqnarray}
from which one deduces, with the asymptotic expansion (\ref{eq:limit}), that
\begin{eqnarray}
\frac{p(\vbar)}{\Delta t}=\frac{w_0}{\cS}\frac{\ii}{\Delta t}+\bigO(\Delta t).
\end{eqnarray}
The real part of this quantity goes to zero as $\tau\to t^-$, as announced.
We thus reach our main result: with the arctan \emph{ansatz} and given any \textit{a priori} time dependence of the core width, the self-force can be written in the \emph{mass form}
\begin{eqnarray}
\label{eq:fesh_screw}
F(t)=2\Re\int_{-\infty}^t\frac{\dd \tau}{t-\tau}
m\left(\vbar(t,\tau)\right)\frac{\dd \vbar}{\dd \tau}(t,\tau)
+2\frac{w_0}{\cS}\frac{v(t)}{a(t)}.
\end{eqnarray}
Thus, the dynamic self-force is expressed exclusively in terms of a mass function already known from stationary calculations. Formerly, Beltz et al.\ (1968) addressed the dynamical problem in such terms, stating that ``the radiation output of an \emph{oscillating} dislocation is entirely determined by the dislocation mass factor $m$''. The present equation shows that this statement can be generalized. However, Eq.\ (\ref{eq:fesh_screw}), while resembling their Eq.\ (37), differs from it in two major ways: the function $\overline{v}(t,\tau)$ enters the integral, rather than the instantaneous velocity $v(\tau)$, and the upper limit of the integral in Eq.\ (\ref{eq:fesh_screw}) does not need any phenomenological cut-off, in close connection to the presence of the rightmost additive term (see Section \ref{sec:counterterm}).

\subsection{Edge dislocation}
The edge case is addressed in the same way. The  appropriate kernel
is given by Eq.\ (\ref{eq:noyauxcg}), with $\Ke_2$ as in (\ref{eq:noyauxcg2}), and $\Ke_1$ as in (\ref{eq:noyauxcg1bis}). To compute the integral over $x$ and $x'$ in the generic expression of $F(t)$,
Eq.\ (\ref{eq:fesh_gen}), use is made of Eq.\ (\ref{eq:firstint}) for the contribution
generated by $\Ke_1$, and of Eq.\ (\ref{eq:secondint}) for that generated by $\Ke_2$. Consider
first the contribution of $\Ke_1$, that comprises four parts. Each of these corresponds to one the four main terms in $\Ke_1(x,t)$, see Eq.\ (\ref{eq:noyauxcg1bis}), which are integrated by appealing, respectively, to integrals (\ref{eq:b3}) (used twice, once with $\cS$ and once with $\cL$), (\ref{eq:b4}) and (\ref{eq:b1}) after elementary changes of variables. The contribution of $\Ke_2$ is integrated using (\ref{eq:b2}). Gathering all contributions, immediate simplifications yield $F(t)$ in the form (\ref{eq:eomexpl}), but now with
\begin{eqnarray}
\label{eq:fevdef}
\hspace{-1em}
\frac{p(v)}{w_0/\cS}\defi4\left(\frac{\cS}{v}\right)^3
\left[\frac{2-(v/\cL)^2}{\sqrt{1-(v/\cL)^2}}-\frac{2-(v/\cS)^2}{\sqrt{1-(v/\cS)^2}}\right]
+\left(\frac{v}{\cS}\right)
\frac{2-(v/\cS)^2}{\left[1-(v/\cS)^2\right]^{3/2}}.
\nonumber\\
\end{eqnarray}
Again, $p=\dd L/\dd v$ with the function $L$ appropriate to edges (Appendix A). The steps leading to Eq.\ (\ref{eq:fesh_screw}) can be reproduced, using the asymptotic expansion (\ref{eq:limit}) to justify the vanishing of the boundary terms in the partial integration. In this way, the self-force for edges is cast in mass form as well.

\subsection{Comments}
\label{sec:comments}
First, it is emphasized that the mass functions $m(v)$ that appear in the problem are \emph{identical} to those given by Hirth, Zbib and Lothe (1998), but with logarithmic factors removed, as discussed in Appendix \ref{eq:remid}. Also, the kernels $K^{\text{e,s}}(x,t)$ in Eqs.\ (\ref{eq:ksalt}b) stand as close relatives to the quasimomentum $p(v)$ in Eqs.\ (\ref{eq:fsvdef}),  (\ref{eq:fevdef}). Indeed,
\begin{eqnarray}
K(x,t)=\frac{\theta(t)}{2w_0} t^{-2} p\left(\frac{x}{t}\right).
\end{eqnarray}
In this writing, $p(x/t)$ must be interpreted as a distribution with, in the edge case, a finite part prescription, in view of Eq.\ (\ref{eq:kealt}).

The structure of $F(t)$ in mass form can be understood as follows. In Eq.\ (\ref{eq:fesh_screw}), the functions $m(v)$ involve terms of the type $(1-v^2/c^2)^{1/2}$, in which the velocity $v$ is compared with a wave velocity $c$ to determine its sub- or supersonic character with respect to $c$. Heuristically ignoring the presence in $\vbar$ of the imaginary contribution of $a$, the replacement in $m(v)$ of $v$ by
\begin{eqnarray}
\vbar(t,\tau)\simeq \frac{\xi(t)-\xi(\tau)}{t-\tau}
\end{eqnarray}
accounts for the fact that whether the dislocation at space-time location $(\xi(t),t)$ goes faster than a wave emitted at $(\tau,\xi(\tau))$ for $\tau<t$, or not, depends on the \emph{average} velocity of the dislocation between these instants rather than on the instantaneous velocity $v(t)$. Of course, $\vbar(t,\tau)$ tends to $v(t)$ in the limit of vanishing time intervals. Moreover (again heuristically), the time-integral in (\ref{eq:fesh_screw}) is tantamount to a sum of ``mass times acceleration'' terms, where the role of the acceleration is played by $2\dd \vbar/\dd \tau$. The factor 2 in the latter is necessary to recover the instantaneous acceleration at small times (Pillon et al., 2007). Indeed, as $\tau\to t^-$ and again ignoring $a$, one has
\begin{eqnarray}
2\frac{\dd \vbar}{\dd \tau}(t,\tau)
\simeq 2\frac{\vbar(t,\tau)-v(\tau)}{t-\tau}=\ddot{\xi}(t)+\frac{2}{3}\dddot{\xi}(t)(\tau-t)+\bigO\left((\tau-t)^2\right).
\end{eqnarray}
The logarithmic integration over past times can be justified from dimensional considerations. Finally, the local term proportional to $v(t)/a(t)$ in the self-force regularizes a divergence in the integral as $a\to 0$ in the relaxation regime at large times, as will be made clear in Section \ref{sec:counterterm}.

\subsection{Link with a result of Eshelby (screw dislocation)}
\label{sec:linkesh}
As recalled in the Introduction, besides proposing his dynamical core equation for screws,
Eshelby (1953) proposed in the same paper an independent derivation of the self-force of a screw, from an electromagnetic analogy  between a moving screw dislocation and a moving current line (his Eq.\ (26) in that reference). He used it as a starting point for further analysis. This equation relies on the arctangent \textit{ansatz} with a \emph{rigid} core $a(t)=a=\text{const}$. Under this constraint, we show that Eq.\ (\ref{eq:fesh_screw}) matches Eshelby's. In our notations, and with $\Delta\xi\defi \xi(t)-\xi(\tau)$ and
$\Delta t\defi t-\tau$, Eshelby's self-force $F(t)$ reads\footnote{Expressions of quite a similar structure were obtained by Lazar (2010), in the context of gradient elasticity.}
\begin{eqnarray}
\hspace{-1em}
\frac{F(t)}{w_0/\cS}=2\Re\int_{-\infty}^t
\dd \tau\, \left\{\frac{\dd v}{\dd \tau}(\tau)+\left[\cS^2-v(\tau)^2\right]
\frac{\partial}{\partial\Delta \xi}\right\}\frac{1}{\sqrt{\cS^2\Delta t^2-(\Delta \xi+\ii a)^2}},\nonumber\\
\label{eq:eshself}
\end{eqnarray}
where we have carried out the integral over Fourier wavevectors in his Eq.\ (26). Using integration by parts, the dislocation being at rest in the remote past,
\begin{eqnarray}
F(t)&=&2\frac{w_0}{\cS}\Re\int_{-\infty}^t \dd \tau\,\left(\frac{\dd }{\dd \tau}\left[\frac{v(\tau)}{\sqrt{\cS^2\Delta t^2-(\Delta \xi+\ii a)^2}}\right]
\right.\nonumber\\
&&\left.{}
+\left\{\left[\cS^2-v(\tau)^2\right]\frac{\partial}{\partial\Delta \xi}-v(\tau)\frac{\dd }{\dd \tau}\right\}\frac{1}{\sqrt{\cS^2\Delta t^2-(\Delta \xi+\ii a)^2}}\right),\nonumber\\
\label{eq:eshtransf}
&=&2\frac{w_0}{\cS}\frac{v(t)}{a}+2w_0\cS\Re\int_{-\infty}^t \dd \tau\,\frac{\Delta \xi+\ii a
-v(\tau)\Delta t}{[\cS^2\Delta t^2-(\Delta \xi+\ii a)^2]^{3/2}},
\end{eqnarray}
where $v(t)/a$ stems from the boundary term at $\tau=t$, and where the remaining term under the integral
has been simplified by evaluating the derivatives. Specializing expression (\ref{eq:vbar})
of $\vbar$ to a rigid core, namely, writing
\begin{eqnarray}
\vbar(t,\tau)=(\Delta \xi/\Delta t)+\ii(a/\Delta t)
\end{eqnarray}
and observing that in this case
$(\dd \vbar/\dd \tau)(t,\tau)=[\vbar(t,\tau)-v(\tau)]/\Delta t$,
Eqs.\ (\ref{eq:fesh_screw}) and (\ref{eq:eshtransf}) are seen to be identical.

\section{Linearized equation of motion}
\label{sec:fdem}
For small velocities, namely, when both $v(t)$ and $\Delta\xi/\Delta t\!=\!\Re\vbar(t,\tau)$ are small compared with $\abar(t,\tau)/\Delta t$ and with the sound velocities, the self-force is linearized by expanding it to first order in these quantities, considered as of the same order. The EoM is then solvable in terms of Fourier or Laplace transforms (Eshelby, 1953; Al'shitz et al., 1971).  Using definition (\ref{eq:vbar}) of $\vbar(t,\tau)$, we assume that the dislocation width depends on time only through the velocity: $a(t)\defi \widetilde{a}(v(t))$. This assumption is discussed in Section \ref{sec:tdcw} where the function $\widetilde{a}(v)$ is determined.

Here and in the rest of this section $a$ stands for $\widetilde{a}(0)$, the core width at rest. To linear order in $\Delta \xi/\Delta t$ and $v$, one has
\begin{eqnarray}
\hspace{-2em}\Re \frac{m(\vbar)}{\Delta t}\frac{\dd \vbar}{\dd \tau}\simeq
\frac{1}{\Delta t }\left[m\left(\ii\frac{a}{\Delta t}\right)\frac{\dd }{\dd \tau}\frac{\Delta\xi}{\Delta t}
+\ii\frac{a}{\Delta t^2}m'\left(\ii\frac{a}{\Delta t}\right)\frac{\Delta\xi}{\Delta t}\right]
=\frac{\dd }{\dd \tau}\left[m\left(\ii\frac{a}{\Delta t}\right)\frac{\Delta\xi}{\Delta t^2}\right]
-m\!\left(\ii\frac{a}{\Delta t}\right)\frac{\Delta\xi}{\Delta t^3},\nonumber\\
&&\label{eq:linexp}
\end{eqnarray}
where, since $m(v)$ essentially depends on $v^2$, $m(\ii a/\Delta t)$ is purely real. This equation makes clear that the time--de\-pen\-den\-ce of $a(t)$ plays no part to linear order. Because the boundary term vanishes upon integrating (\ref{eq:linexp}) over $\tau$ the linearized self-force reads
\begin{eqnarray}
F^{\text{lin}}(t)&\defi&2\frac{w_0}{a}\frac{v(t)}{\cS}
-2\int_{-\infty}^t\dd \tau\, m\!\left(\ii\frac{a}{\Delta t}\right)\frac{\Delta\xi}{\Delta t^3}\nonumber\\
&=&2\frac{w_0}{a\cS}\dot{\xi}(t)-2\xi(t)\int_{-\infty}^t\frac{\dd \tau}{\Delta t^3} m\!\left(\ii\frac{a}{\Delta t}\right)+2\int_{-\infty}^t\frac{\dd \tau}{\Delta t^3} m\!\left(\ii\frac{a}{\Delta t}\right)\xi(\tau)\nonumber\\
&=&2\frac{w_0}{a\cS}\dot{\xi}(t)+\frac{2}{a^2}[W(\ii\infty)-W(0)]\xi(t)
+2\int_{-\infty}^t\frac{\dd \tau}{\Delta t^3} m\!\left(\ii\frac{a}{\Delta t}\right)\xi(\tau)\nonumber\\
\label{eq:flinexpl}
&=&2\frac{w_0}{a\cS}\dot{\xi}(t)-2\frac{W(0)}{a^2}\xi(t)
+2\int_{-\infty}^t\frac{\dd \tau}{\Delta t^3} m\!\left(\ii\frac{a}{\Delta t}\right)\xi(\tau).
\end{eqnarray}
In going to the third line, a change of variable $u=i a/\Delta t$ has been used, together with the identity $m(v)=W'(v)/v$ between $m(v)$ and the total line energy $W(v)$. This relationship is discussed in Appendix A where $W(0)$ is given and where it is shown that $W(\ii\infty)=0$. The result is of the form
\begin{eqnarray}
F^{\text{lin}}(t)=-C_0\xi(t)+C_1\dot{\xi}(t)+\int_{-\infty}^{+\infty} \dd \tau\, C_2(t-\tau)\xi(\tau),
\end{eqnarray}
where $C_0$, $C_1$ are constants, and where $C_2$ is a causal response kernel with $C_2(t)=0$ if $t<0$.
Fourier-transforming $F^{\text{lin}}(t)$ with respect to time yields the damped linear oscillator form (Nabarro, 1951)
\begin{eqnarray}
\label{eq:fomega}
F^{\text{lin}}(\omega)\defi [-\omega^2 \wm(\omega)-\ii\omega\,\walpha(\omega)]\xi(\omega),
\end{eqnarray}
where $\wm(\omega)=[C_0-\Re C_2(\omega)]/\omega^2$ is the frequency-dependent mass and $\walpha(\omega)=C_1+\Im C_2(\omega)/\omega$ is the frequency-dependent damping coefficient, two real quantities for $\omega$ real. The tilde distinguishes them from the previously introduced mass $m(v)$ and phenomenological damping constant $\alpha$.

For the screw, $\wm(\omega)$ and $\walpha(\omega)$ were exploited by Al'shitz et al.\ (1971) on the basis of Eshelby's linearized self-force, Eq.\ (28) in (Eshelby, 1953). They were revisited by Pillon et al.\ (2007) who spotted an incorrect $1/2$ factor in the linearized expression,\footnote{This explains the discrepancy between Eq.\ (\ref{eq:mlaps}) below, and Eshelby's (1953) Eq.\ (30). }and started from Eshelby's Eq.\ (26) instead. Here, this amounts to proceeding from the equivalent Eqs.\ (\ref{eq:fesh_screw}) or (\ref{eq:eshtransf}). Introduce characteristic times $\tSL\defi a/\cSL$, and the reference mass
\begin{eqnarray}
m_0\defi w_0/\cS^2=\mu b^2/(4\pi\cS^2).
\end{eqnarray}
Equation (3) in Pillon et al.\ (2007), namely,
\begin{eqnarray}
\label{eq:pillon}
\frac{F^{\text{lin}}(t)}{m_0}=-2 \frac{\xi(t)}{\tS^2}+2\frac{\dot{\xi}(t)}{\tS}+2\int_{-\infty}^t \dd \tau\, \frac{\xi(\tau)}{[(t-\tau)^2+\tS^2]^{3/2}},
\end{eqnarray}
is nothing but the particularization of Eq.\ (\ref{eq:flinexpl}) to screw dislocations. Let $I_n$, $K_n$ and $\mathbf{L}_n$ denote the modified Bessel and Struve function  (e.g., Abramowitz and Stegun, 1972), and introduce for convenience auxiliary functions $I\!\mathbf{L}_\nu$ as
\begin{eqnarray}
I\!\mathbf{L}_\nu(z)\defi \frac{\pi}{2}\left[I_\nu(z)-\mathbf{L}_\nu(z)\right]\qquad (\nu=0,1,2).
\end{eqnarray}
The following expressions were obtained by Pillon et al.\ (2007) from Eq.\ (\ref{eq:pillon}):
\begin{subequations}
\label{eq:malwscrew}
\begin{eqnarray}
\label{eq:mws}
\frac{\wm(\omega)}{m_0}&=&\frac{2}{\omega^2\tS^2}\bigl[1-|\omega|\tS K_1(|\omega|\tS)\bigr]\\
\label{eq:mwslow}
&\simeq& -\log\left(\frac{|\omega|\tS}{2\ee^{\frac{1}{2}-\gamma}}\right)+\bigO(\omega^2\log|\omega|),\\
\label{eq:mwshigh}
&\sim& 2 (\tS\omega)^{-2}+\text{exponentially decaying terms},
\end{eqnarray}
\begin{eqnarray}
\label{eq:alps}
\tS\frac{\walpha(\omega)}{m_0}&=&2\left\{1+\frac{\pi}{2}
\bigl[I_1(|\omega|\tS)-\mathbf{L}_{-1}(|\omega|\tS)\bigr]\right\}=2 I\!\mathbf{L}_1(|\omega|\tS)\\
\label{eq:alpslow}
&\simeq& \frac{\pi}{2}|\omega|\tS+\bigO(\omega^2)\\
\label{eq:alpshigh}
&\sim& 2 +\bigO(\omega^{-2})
\end{eqnarray}
\end{subequations}
($\gamma$ is Euler's constant). The identity $\mathbf{L}_1(z)=\mathbf{L}_{-1}(z)-2/\pi$ has been used in Eq.\ (\ref{eq:alps}) to further reduce $\walpha(\omega)$. Equations (\ref{eq:mws}) and (\ref{eq:alps}) are readily recovered upon Fourier transforming  Eq.\ (\ref{eq:flinexpl}) with  Eq.\ (\ref{eq:ws}) by means of integral (\ref{eq:c2}).

To our knowledge, the frequency-dependent response coefficients of the edge are not available (see however Kiusalaas and Mura, 1964b). The Fourier transform of Eq.\ (\ref{eq:flinexpl}) combined with (\ref{eq:we}) is carried out with the help of integrals (\ref{eq:c1})--(\ref{eq:c4}). After some straightforward simplifications involving the identities $K_0(x)=-2K_1(x)/x+K_2(x)$, $I_0(x)=2I_1(x)/x+I_2(x)$ and $\mathbf{L}_0(x)=2\mathbf{L}_1(x)/x+\mathbf{L}_2(x)+2 x/(3\pi)$ (Abramowitz and Stegun, 1972) one arrives at:
\begin{subequations}
\label{eq:malwedge}
\begin{eqnarray}
\frac{\wm(\omega)}{m_0}
&=&\frac{4}{\omega^2\tS^2}\Bigl\{\bigl[1+2|\omega|\tS
K_{1}(|\omega|\tS)+6 K_2(|\omega|\tS)\bigr]\nonumber\\
\label{eq:mwe}
&&{}-\frac{\tL^2}{\tS^2}\bigl[1+2|\omega|\tL
K_{1}(|\omega|\tL)+6 K_2(|\omega|\tL)\bigr]\Bigr\}+2K_0(|\omega|\tS)\\
\label{eq:mwelow}
&\simeq&-\log\left[\left(\frac{|\omega|\tS}{2\ee^{\frac{1}{4}-\gamma}}\right)\left(\frac{|\omega|\tL}{2\ee^{-\frac{1}{4}-\gamma}}\right)^{\left(\tL/\tS\right)^4}
\right]+\bigO(\omega^2\log|\omega|),\\
\label{eq:mwehigh}
&\sim& 4\left(1-\frac{\tL^2}{\tS^2}\right)(\tS\omega)^{-2}+\text{exponentially decaying terms},
\end{eqnarray}
\begin{eqnarray}
\tS\frac{\walpha(\omega)}{m_0}&=&
\frac{4}{|\omega|\tS}\Bigl\{
\bigl[6 I\!\mathbf{L}_2(|\omega|\tS)-2|\omega|\tS\,I\!\mathbf{L}_1(|\omega|\tS)\bigr]\nonumber\\
\label{eq:alpe}
&&-\frac{\tL^2}{\tS^2}
\bigl[6 I\!\mathbf{L}_2(|\omega|\tL)-2|\omega|\tL\,I\!\mathbf{L}_1(|\omega|\tL)\bigr]
\Bigr\}+2|\omega|\tS I\!\mathbf{L}_0(|\omega|\tS)\\
\label{eq:alpelow}
&\simeq& \frac{\pi}{2}\left(1+\frac{\tL^4}{\tS^4}\right)|\omega|\tS+\bigO(\omega^2)\\
\label{eq:alpehigh}
&\sim& 2 +\bigO(\omega^{-2}).
\end{eqnarray}
\end{subequations}
Equation (\ref{eq:mwelow}) is rewritten as
\begin{eqnarray}
\label{eq:mwe2}
\wm(\omega)\simeq-m(0)\log(|\omega|t^*), \qquad t^*=\frac{e^\gamma}{2}\left( e^{-\frac{1}{4}}\tS\right)^{\frac{1}{1+\phi}}\left(e^{\frac{1}{4}}\tL\right)^{\frac{\phi}{1+\phi}},
\end{eqnarray}
where $\phi\defi (\tL/\tS)^4$ and  $m(0)=(1+\phi)m_0$ is the rest mass factor of the edge.

\begin{figure}
\centering
\includegraphics[width=16cm]{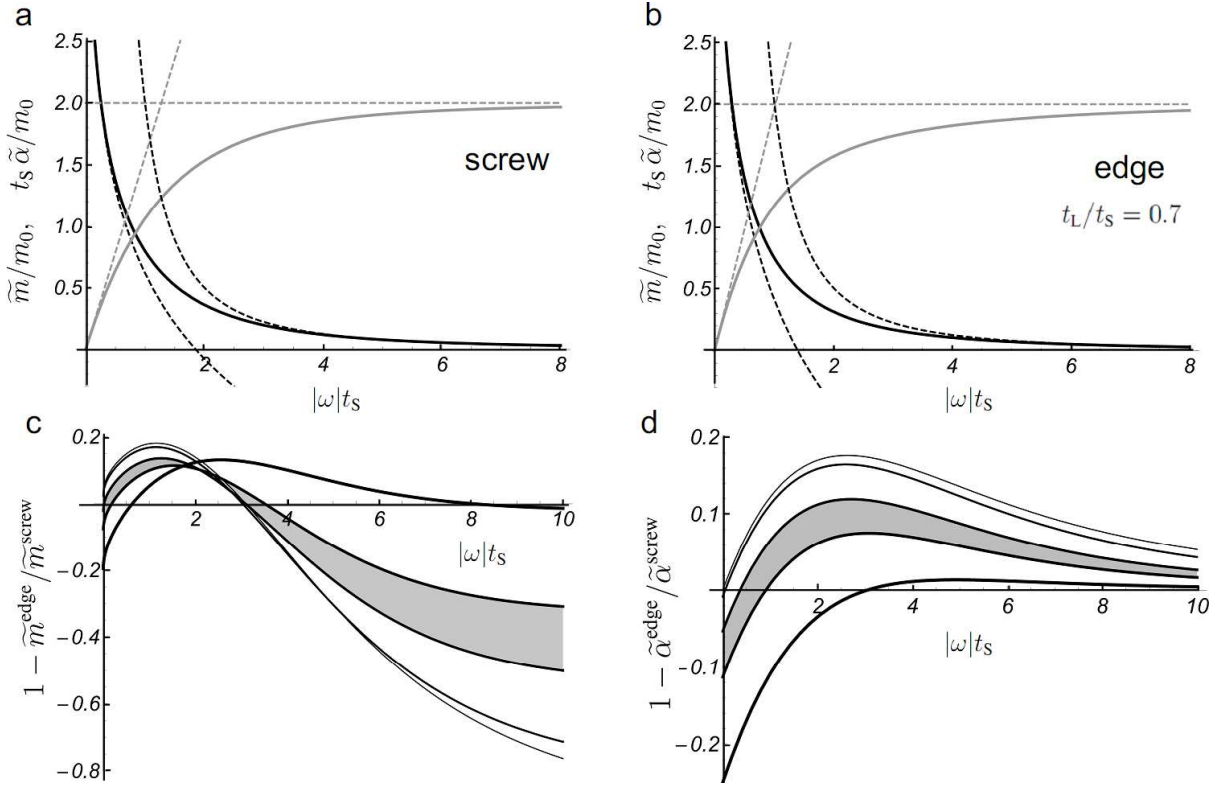}
\caption{\label{fig:fig1} Top: normalized frequency-dependent effective mass $\widetilde{m}$ (solid, black) and damping coefficient $\widetilde{\alpha}$ (solid, grey) for (a) screw, and (b) edge dislocation for $\tL/\tS=0.7$. The low-frequency and asymptotic behaviors expressed by Eqs.\ (\ref{eq:mwslow}), (\ref{eq:mwshigh}), (\ref{eq:alpslow}), and (\ref{eq:alpshigh}) for the screw, and corresponding equations for the edge, are dashed. Bottom: relative difference  between screw and edge for (c) $\widetilde{m}(\omega)$ and (d) $\widetilde{\alpha}(\omega)$, plotted vs.\ frequency for $\tL/\tS=0.0$, $0.3$, $0.48$, $0.58$ and $0.7$ (line thickness increases with $\tL/\tS$). The range of interest for most metals ($0.48\leq \tL/\tS\leq 0.58$) is shaded.}
\end{figure}

By employing Laplace transforms (LT) of variable $s$ instead of Fourier transforms, Eq.\ (\ref{eq:fomega}) becomes $F^{\text{lin}}(s)\defi s^2 \wwm(s)\xi(s)$. In view of the relationship between FTs and LTs the function $\wwm$ is such that $\wwm(-\ii\omega)=\wm(\omega)+\ii\walpha(\omega)/\omega$. The FTs are then deduced from
\begin{subequations}
\label{eq:laplace}
\begin{eqnarray}
\label{eq:mlaps}
\frac{\wwm(s)}{m_0}&=&\frac{2}{(s\tS)^2}\left[s\tS\,\HY_1(s\tS)-1\right]\qquad \text{(screw)},\\
\label{eq:mlape}
                   &=&\frac{4}{(s\tS)^2}\Bigl\{
\bigl[6 \HY_2(s\tS)-2 s\tS\,\HY_1(s\tS)-1\bigr]\nonumber\\
&&-\frac{\tL^2}{\tS^2}
\bigl[6 \HY_2(s\tL)-2 s\tL\,\HY_1(s\tL)-1\bigr]\Bigr\}+2 \HY_0(s\tS)\quad \text{(edge)},
\end{eqnarray}
\end{subequations}
where the auxiliary function $\HY_\nu(z)\defi\frac{\pi}{2}\left[\mathbf{H}_\nu(z)-Y_\nu(z)\right]$ is defined in terms of the Struve function $\mathbf{H}_\nu$ and Bessel function of the second kind $Y_\nu$ (Abramowitz and Stegun, 1972). The function $\wwm(s)$ is free of singularities in the complex plane, except for a cut on the negative real axis. It will be used in Section \ref{sec:asym}.

The functions $\widetilde{m}$ and $\widetilde{\alpha}$ are drawn in Figs.\ \ref{fig:fig1}(a), (b) for $\tL/\tS\simeq 0.70$, i.e., $\cL\simeq 1.4\, \cS$, a situation where the screw and the edge behave in quite a similar way. The damping coefficient $\widetilde{\alpha}(\omega)$ is always positive. So is $\widetilde{m}^{\text{screw}}(\omega)$. The ratio $\tL/\tS=\cS/\cL=[2(1-\nu)/(1-2\nu)]^{-1/2}$ decreases from $1/\sqrt{2}$ to $0$ when the Poisson ratio $\nu$ increases from $0$ to $1/2$, so that $\cS<\cL/\sqrt{2}$. In this range, $\widetilde{m}^{\text{edge}}(\omega)>0$, as illustrated by the asymptotic value in Eq.\ (\ref{eq:mwehigh}), which changes sign at $\cS=\cL$. For metals $\nu$ varies between $0.032$ (beryllium, $\tL/\tS\simeq 0.69$) and $0.45$ (thallium $\tL/\tS\simeq 0.30$), while it stands in the range $0.25$--$0.35$ for most of them (Lide, 2005). This corresponds to $\tL/\tS\sim 0.48$--$0.58$. Comparisons between screw and edge for $\widetilde{m}$ and $\widetilde{\alpha}$ are displayed in Figs.\ \ref{fig:fig1}(c) and \ref{fig:fig1}(d) for $0\leq \tL/\tS\leq 0.7$. Positive values indicate that the screw dominates over the edge. While various situations are met for $\widetilde{m}(\omega)$, depending on the frequency range and on the Poisson ratio, edge dislocations experience in general more radiative damping than screws at low frequencies. The situation is reversed at high frequencies. These observations remain somewhat formal, in view of our using planar cores and isotropic elasticity.

\section{The local term as a counter-term}
\label{sec:counterterm}
The purpose of this section is to show that the local term in Eq.\ (\ref{eq:fesh_screw}), inversely
proportional to the dislocation width $a$, acts as a
\emph{counter-term}\footnote{The denomination is borrowed from quantum electrodynamics.} (CT), which compensates for one specific contribution to the self-force integral that explodes at finite times as $a^{-1}$ in the Volterra limit $a\to 0$. This property relates to the observations made in Section \ref{sec:lossterm} while computing the stress $\sigma_\eta$ of a Volterra screw.

Although we derived it for an isotropic medium, let us assume that the expression of the self-force in its mass form is of general validity, and consider as known the total energy function $W(v)$ (see Appendix A). The game to be played is to pretend that that the local contribution is unknown. Writing it as $\text{CT}$, the self-force reads
\begin{eqnarray}
\label{eq:sf}
F(t)=\text{CT}+2\Re\int_{-\infty}^t\frac{\dd \tau}{t-\tau}
m\left(\vbar(t,\tau)\right)\frac{\dd \vbar}{\dd \tau}(t,\tau).
\end{eqnarray}
The unknown CT stems from requiring that an expansion of $F(t)$ in powers of $a$ does not lead to infinities as the stationary limit is approached \emph{in the subsonic regime}.

Indeed, let the dislocation undergo at $t=0$ an instantaneous velocity jump between arbitrary initial and final velocities $\vi$ and $\vf$, while it passes at the origin, so that $\xi(0)\defi 0$. We use the symbol $\jump{f}\defi f_{\rm f}-f_{\rm i}$ to denote the jump of any quantity $f$ between some initial and some final state. We take $v(t)=\vi\,\theta(-t)+\vf\,\theta(t)$, and suppose first for simplicity that $a$ is constant. This assumption is relaxed in the next sections. Then, with $\jump{v}=\vf-\vi$ and for positive times,
\begin{eqnarray}
\vbar(t,\tau)=\left\{
\begin{array}{cc}
\vi+\frac{\jump{v}\,t+\ii a}{t-\tau}\qquad&\text{if}\qquad \tau\leq 0,\\
\vf+\frac{\ii a}{t-\tau}\qquad&\text{if}\qquad 0\leq\tau<t.
\end{array}
\right.
\end{eqnarray}
We carry out the change of variables $u=\vbar(t,\tau)$, such that
${{\rm d}u}=({\rm d}\vbar/{\rm d}\tau){\rm d}\tau$. Thus,
\begin{eqnarray}
\frac{1}{t-\tau}=\left\{
\begin{array}{cc}
\frac{u-\vi}{\jump{v} t+\ii a}&\text{if }\tau\in(-\infty,0],\quad\text{ i.e., for
$u$ on a path }[\vi,\vf+\ii(a/t)];\nonumber\\
\frac{u-\vf}{\ii a}&\text{if }\tau\in[0,t],\quad\text{ i.e., for
$u$ on a path }[\vf+\ii(a/t),+\ii\infty).
\end{array}
\right.
\\
\end{eqnarray}
Using identity (\ref{eq:massident2}), this leads to the following expression of the self-force:
\begin{eqnarray}
F-\text{CT}&=&
2\Re\Biggl[\int_{\vi}^{\vf+\ii\frac{a}{t}}\frac{{\rm d}u(u-\vi)}{(\jump{v}t+\ii a)u}\frac{{\rm d}W}{{\rm d}u}(u)
+\int_{\vf+\ii\frac{a}{t}}^{\ii\infty} {\rm d}u\,
\frac{u-\vf}{\ii a u}\frac{{\rm d}W}{{\rm d}u}(u)\Biggr]
\nonumber\\
&=&2\Re
\Biggl[\frac{W\bigl(\vf+\ii\frac{a}{t}\bigr)-W(\vi)}{\jump{v} t+\ii a}
-\frac{\vi}{\jump{v} t+\ii a}\int_{\vi}^{\vf+\ii\frac{a}{t}}\frac{{\rm d}u}{u}\frac{{\rm d}W}{{\rm d}u}(u)
\nonumber\\
&&\quad{}+\frac{W(\ii\infty)
-W\bigl(\vf+\ii\frac{a}{t}\bigr)}{\ii a}
-\frac{\vf}{\ii a}\int_{v+\ii\frac{a}{t}}^{\ii\infty}
\frac{{\rm d}u}{u}\frac{{\rm d}W}{{\rm d}u}(u)\Biggr].
\end{eqnarray}
The remaining integrals are readily evaluated in terms of the kinetic energy function of the field, $\Wk$, thanks to Eq.\ (\ref{eq:identwkw}), and further simplified in terms of the ``stationary Lagrangian'' $L(v)$  (\ref{eq:lagdef}), and of the momentum relationship $p(v)=2\Wk(v)/v$, see Eq.\ (\ref{eq:beltz}). One ends up with
\begin{subequations}
\begin{eqnarray}
\label{eq:fself}
F&=&\Delta\text{CT}+\frac{2}{a}\Im\frac{\jump{v} L\left(\vf+\ii \frac{a}{t}\right)+\ii \frac{a}{t}L(\vi)}{\jump{v}+\ii \frac{a}{t}}\\
\Delta\text{CT}&\defi&\text{CT}-\frac{2}{a}\Im
\left[\vf p(u)-W(u)\right]_{u=\ii\infty}.
\end{eqnarray}
\end{subequations}
The self-force is now straightforwardly expanded to leading orders in powers of $a/t$, and simplified using $\dd L/\dd v=p$. This expansion, which probes the relaxation regime after the jump, is relevant also in the Volterra limit $a\to 0$. It reads
\begin{eqnarray}
\label{eq:fselfexpan}
F=\Delta\text{CT}+\frac{2}{a}\Im L(\vf)+\frac{2}{t}\Re\left[p(\vf)-\frac{\jump{L}}{\jump{v}}\right]+\bigO(a).
\end{eqnarray}
For it to provide a finite leading-order term for $|v_{\rm i,f}|<\cS$ when all energies are real, it is necessary that $\Delta\text{CT}=0$, from which the counter-term must be
\begin{eqnarray}
\label{eq:ctv}
\text{CT}=\frac{2}{a}\Im
\left[v(t)p(\ii\infty)-W(\ii\infty)\right],
\end{eqnarray}
where $v(t)$ stands for $\vf$ since $t>0$. It is shown in Appendix A that, for both screw and edge, $W(i\infty)=0$ and $p(\ii\infty)=2[\Wk(u)/u]_{u=\ii\infty}=\ii w_0/\cS$. Thus, the local term of EoM  is retrieved exactly, namely, $\text{CT}=2 w_0 v(t)/(a\cS)$. This calculation suggests that its general expression is
\begin{eqnarray}
\label{eq:ctv2}
\text{CT}=2\Im[p(\ii\infty)]\frac{v(t)}{a(t)}.
\end{eqnarray}
Eventually, the asymptotic self-force is given by Eq.\ (\ref{eq:fselfexpan}) with $\Delta\text{CT}=0$, which is implied in the rest of the paper when referring to (\ref{eq:fselfexpan}). The way it is obtained indicates that $v_{\rm i,f}$ should be interpreted as $v_{\rm i,f}+\ii 0^+$, where the infinitesimal positive imaginary part is required to get correct determinations of the energies above $\cS$.

For velocities $|\vf|>\cS$, the term $(2/a)\Im L(\vf+\ii 0^+)$ in Eq.\ (\ref{eq:fselfexpan}) is non-zero and represents the drag force due to concentrated radiation at Mach fronts (Weertman, 1969), to be counter-balanced in steady motion by the applied force (e.g., Rosakis, 2001). Its blowing-up as $a\to 0$ -- due to a constant amount of energy on an infinitely thin front -- together with that of the phenomenological drag force $F_{\text{D}}$ (\ref{eq:dragforce}), suggests that any kind of instantaneous drag stress not proportional to the dislocation width or to higher powers thereof leads in the limit to an infinite self-force contribution, consistently with the presence of the Dirac term in Eq.\ (\ref{eq:stressy2}).

Moreover, bearing in mind that $p=\dd L/\dd v$, the quantity $\jump{L}/\jump{v}$ represents the quasimomentum of the dislocation \emph{at} jump time, which stands as the relevant `initial' impulsion prior to relaxation. Consequently, $p(\vf)-\jump{L}/\jump{v}$ in Eq.\ (\ref{eq:fselfexpan}) is the remainder of the impulsion transfer. We emphasize that Lagrangians with imaginary parts are only rarely encountered in classical mechanics, one noteworthy instance being in Dekker's (1975) approach to dissipative systems by a complex Lagrangian.

Equation (\ref{eq:fselfexpan}) embodies former results by Clifon and Markenscoff (1981) for subsonic dislocations jumping instantaneously  from rest ($\vi=0$) to a velocity $v=\vf$. Indeed, since $\Wk(0)=0$, one has $L(0)=-W(0)$ and $L(v)-L(0)=v p(v)-[W(v)-W(0)]$, so that Eq.\ (\ref{eq:fselfexpan}) reduces to
\begin{eqnarray}
\label{eq:cmforce_sub}
F(t)&=&2 K(v)/ (v t),
\end{eqnarray}
where $K(v)\defi W(v)-W(0)$ is the kinetic energy (\ref{eq:Kvm}) of the stationary dislocation. By using expressions (\ref{eq:ws}) and (\ref{eq:we}) of $W(v)$ in Eq.\ (\ref{eq:cmforce_sub}) and reorganizing terms, one sees that Eq.\ (\ref{eq:cmforce_sub}) matches Eqs.\ (40) and (42) (screws), and (47) and (49) (edges) in the above reference. Wu (2002) provides a generalization of Eq.\ (\ref{eq:cmforce_sub}) to anisotropic media.

Finally, it can be instructive to retrieve this Eq.\ (\ref{eq:cmforce_sub}) directly from Eq.\ (\ref{eq:stressy2}) (screw case). Let $x=vt+\epsilon$ in $\sigma_\eta(x,t)$, and carry out a power expansion of in $\epsilon$. The first-order term is proportional to $1/\epsilon$ and represents a divergent contribution to be ignored, thanks to the principal value prescription. The next term is a finite constant. Multiplying it by $\rho(x,t)=-b\delta(x-vt)$ and integrating over $x$ yields Eq.\ (\ref{eq:cmforce_sub}) back.

\section{Stationary approximation for the time-dependent core width}
\label{sec:tdcw}
Before going on with dynamical regime changes, we must discuss the time-dep\-end\-ence of the core width. To obtain the EoM from the core equation, a ``projection'' of the latter on $\rho(x,t)$ was used in Section \ref{sec:eomprinc}, conjointly with an arctangent \textit{ansatz} of variable width $a(t)$. Although this would be required to complete the theory, we shall not attempt to derive an evolution equation for  $a(t)$ by like means. Instead, we continue to assume an instantaneous dependence $a(t)\defi\widetilde{a}(v(t))$, written as $a(v)$ for brevity. Some limitations of this assumption are however pointed out.

This assumption is the ``minimal'' one leading to a meaningful stationary regime. On the one hand ``relativistic'' core distortion evidently imposes some well-defined dependence on velocity; on the other hand, the issue of the stationary EoM has been examined by Rosakis (2001), who pointed out that the stationary PN equation being conservative, stress-driven subsonic stationary motion is impossible without adding in some phenomenological friction term. This resulted in his \emph{Model I}, already referred to in Section \ref{sec:overview}, which fully determines $a(v)$, friction included.

The friction stress (\ref{eq:dragstress}) gives rise  to the drag force $\FD$ (\ref{eq:dragforce}) in the EoM (\ref{eq:eomgen}). The latter is consistent with \emph{Model I} at stationarity: the stationary self-force results from letting $t\to\infty$ in Eq.\ (\ref{eq:fself}) or more simply in Eq.\ (\ref{eq:fselfexpan}), which is also relevant to large-time behavior. Accounting for $\FD$ yields the stationary stress/velocity relationship
\begin{eqnarray}
\label{eq:eomros1}
4\alpha \frac{w_0}{a(v)}\frac{v}{\cS}+\frac{2}{a(v)}\Im L(v+\ii 0^+)=b\siga.
\end{eqnarray}
To make the connection with \emph{Model I} conspicuous, introduce the \emph{theoretical shear strength} $\sigma_{\text{th}}\defi\mu b/(2\pi d)$ (Hirth and Lothe, 1982) where $d$ is the interatomic plane separation, so that $2 w_0=d b\,\sigma_{\text{th}}$. Introduce next functions $C(v)$ (complex), and $A(v)$ and $B_\alpha(v)$ (real) defined as
\begin{eqnarray}
\label{eq:defABC}
C(v)\defi -A(v)+\ii B_\alpha(v)\defi \frac{1}{2 w_0}L(v+\ii 0^+)+\ii\alpha\frac{v}{\cS},
\end{eqnarray}
in terms of which Eq.\ (\ref{eq:eomros1}) is rewritten as $2[d/a(v)]B_\alpha(v)=\siga/\sigma_{\text{th}}$.
Introduce moreover the modulus $D(v)\defi |C(v)|=[A^2(v)+B^2_\alpha(v)]^{1/2}$. Finally, define a function $B(v)\defi B_0(v)$ for $\alpha=0$, such that $B_\alpha(v)=B(v)+\alpha(v/\cS)$. For $c>0$, and with the principal determination of the square root, one has for real $v$, as $\epsilon\to 0^+$,
\begin{eqnarray}
\label{eq:princdet}
\sqrt{1-(v/c+\ii\epsilon)^2}=\sqrt{|1-v^2/c^2|}\,\bigl[\theta(c-|v|)-\ii\sign(v)\theta(|v|-c)\bigr].
\end{eqnarray}
From this, and by specializing $L(v)$ to screw and edge dislocations using Eqs.\ (\ref{eq:lagse}ab), one verifies that the functions $A(v)$, $B(v)$, $B_\alpha(v)$ and $D(v)$ in Eq.\ (\ref{eq:defABC}) are exactly those in (Rosakis, 2001), with $A(v)$ and $B(v)$ acting as coefficients of the nonlocal and local terms in the augmented Weertman equation of \emph{Model I}. The function $A(v)$ is nonzero only for $|v|<\cS$ (screws) or $|v|<\cL$ (edges), whereas $B(v)$ is non-zero only for $|v|>\cS$ (screws and edges).

In this framework, the function $a(v)$ is readily obtained. As mentioned in Section \ref{sec:overview}, the dynamical core equation (\ref{eq:pndyn}) reduces to \emph{Model I} at stationarity for a sine pullback force $f'(\eta)=\sigma_{\text{th}}\sin(2\pi\eta/b)$. Then, a necessary condition for the arctan ansatz (\ref{eq:etaansatz}) to be a solution of the latter is that $a(v)\defi 2 d D(v)$, where the above-defined $D$ function is proportional to the energy dissipation rate  (Rosakis, 2001). This ``external'' argument is our present substitute for the missing evolution equation for $a(t)$. With this $a(v)$, Eq.\ (\ref{eq:eomros1}) reduces to Rosakis's kinetic equation:
\begin{eqnarray}
\label{eq:eomros2}
B_\alpha(v)/D(v)=\siga/\sigma_{\text{th}}\qquad (\leq 1).
\end{eqnarray}
The inequality stems from the relationship between $A$, $B_\alpha$ and $D$ and indicates the breakdown of the single-dislocation solution when $\siga>\sigma_{\rm th}$.\footnote{With drag term added, Weertman's equation admits then a ``staircase-like'' solution, which represents a train of kinematically-nucleated supersonic dislocations. Briefly evoked by Eshelby (1956), it has been explicitly given by Movchan, Bullough and Willis (1998) in a different context.} From the above, the core width and the kinetic relation are expressed in terms of the Lagrangian and parameter $\alpha$ in a succinct (but equivalent) formulation alternative to Rosakis's,
\begin{subequations}
\label{eq:asig}
\begin{eqnarray}
\label{eq:av}
a(v)&=&2d \left|L(v+\ii 0^+)/(2 w_0)+\ii\alpha (v/\cS)\right|,\\
\siga&=&\sigma(v)\defi\sigma_{\rm th}\sin\mathop{\rm Arg} \left[L(v+\ii 0^+)/(2 w_0)+\ii\alpha(v/\cS)\right],
\end{eqnarray}
\end{subequations}
with, in particular, $a(0)$ equal to $d$ (screw) or $2d[1-(\cS/\cL)^2]=d/(1-\nu)$ (edge).

\begin{figure}
\centering
\includegraphics[width=18cm]{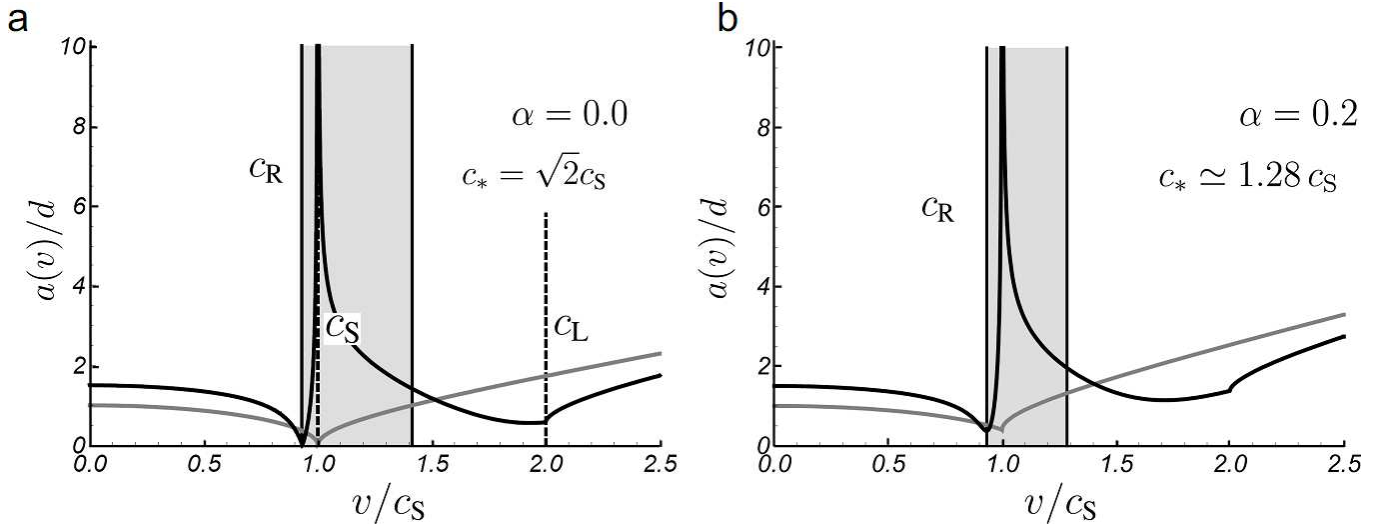}
\caption{\label{fig:fig2} Core width $a(v)$ vs.\ dislocation velocity $v$ in Rosakis's \emph{Model 1}, Eq.\ (\ref{eq:av}), for $\alpha=0$ and $\alpha=0.2$. Black: edge dislocation; grey: screw dislocation. Shaded zone: $\alpha$-dependent instability domain $v\in(\cR,c_*)$. A value $\cL=2\cS$ is used.}
\end{figure}

Graphs for $\sigma(v)$ are given by Rosakis (2001). The supersonic regime necessarily takes place at saturated stress value $\sigma(v)=\sigma_{\text{th}}$.\footnote{This unrealistic feature of \emph{Model I} is alleviated in Rosakis's \emph{Model II} with gradient (2001), which admits larger stresses. No such correction is attempted here.}  Fig.\ \ref{fig:fig2} represents the core width $a(v)$, Eq.\ (\ref{eq:av}), for drag coefficients $\alpha=0$ and $\alpha=0.2$ (see also Fig.\ 1 in Pillon et al., 2007). In the absence of drag, $a(v)$ vanishes either at $v=\cS$ (screw) or at the Rayleigh wave velocity $\cR\simeq 0.93 \cS$ (edge), $\cR$ being the solution $v$ of $A(v)=0$ (Eshelby, 1949). With drag added, $a(v)$ is everywhere strictly positive.

Fig.\ \ref{fig:fig2} illustrates the fact that the edge core width blows up at $v=\cS$, which dramatically reduces the radiative drag, and points towards some pathology of the arctangent \textit{ansatz}. This blowing-up probably constitutes the signature of a rich dynamical process involving core dissociation into several partials, followed by recombination as $\cS$ is overcome, as reported in some molecular dynamics simulations (Li and Shi, 2002; Olmsted et al., 2005). Obviously, a single-dislocation \textit{ansatz} cannot be much accurate in trying to capture such an event, but the mere existence of some analytical signature is worth mentioning.

However, within \emph{Model I} for an edge, an instability range $\cR<|v|<c_*$ exists, identified by a negative derivative $\partial\sigma/\partial v<0$, where $\cS<c_*<\cL$ is the solution of $\partial\sigma(v=c_*)/\partial v=0$ (Rosakis, 2001). The upper bound $c_*$ depends on $\alpha$, and decreases from the ``radiation-free'' velocity $c_*=\sqrt{2}\cS$ (Eshelby, 1949; Weertman, 1969; Gumbsch and Gao, 1999; Gao et al., 1999) for $\alpha=0$ down to $\cS$ as $\alpha$ increases. No steady-state is possible for such velocities, which probably makes inadequate the approximation $a(t)=a(v(t))$ there. Since this range overlaps the interval $[\cR,\cL]$, and for lack of a dynamical governing equation for $a(t)$, we shall only consider velocities less than $\cR$ in numerical applications. This issue concerns $a(t)$, but not the foregoing calculations.

\section{Velocity changes with varying core width}
\label{sec:generalization}
\subsection{Simultaneous jump in velocity and core width}
\label{sec:jump}
The calculation of Section \ref{sec:counterterm} extends to negative times and to the case of a non-constant $a$ by going back to definition (\ref{eq:vbar}) of $\vbar(t,\tau)$. Introduce initial and final values of $a(t)$ as $\aif$, and $\abar\defi(\ai+\af)/2$. When $t<0$, $a(t)\equiv\ai$ and the previous method straightforwardly provides
\begin{eqnarray}
\label{eq:part1}
F(t)&=&\frac{2}{\ai}\Im L(\vi+\ii 0^+)\quad\text{if}\quad t<0.
\end{eqnarray}
For $t>0$, the jump of $a(\tau)$ at $\tau=0$ creates a difficulty, in view of the derivative
\begin{eqnarray}
\label{eq:derivvbar}
\frac{\dd \vbar}{\dd \tau}(t,\tau)
=\frac{\vbar(t,\tau)-v(\tau)}{t-\tau}+\frac{\ii \dot{a}(\tau)}{2(t-\tau)}
=\frac{\vbar(t,\tau)-v(\tau)}{t-\tau}+\frac{\ii}{2 t}\dot{a}(\tau).
\end{eqnarray}
Indeed, the discontinuity implies that $\dot{a}(\tau)\propto \delta(\tau)$, which we have used in simplifying the last term of (\ref{eq:derivvbar}). Thus, a separate treatment of the vicinity of $\tau=0$ in the integral $\int_{-\infty}^t\dd\tau$ that defines $F(t)$ in Eq.\ (\ref{eq:fesh_screw}) is required. We write its contribution as ($t>0$)
\begin{eqnarray}
F_{\tau=0}(t)&\defi&2\Re\lim_{\epsilon\to 0}\int_{-\epsilon}^\epsilon \frac{\dd\tau}{t-\tau}m(\vbar(t,\tau))\frac{\ii}{2 t}\dot{a}(\tau)\nonumber\\
\label{eq:fjump}
&\approx&\frac{2}{t}\Re\int_{\ai}^{\af}m\left(\vf+\frac{\ii}{2 t}(\af+a)\right)\frac{\ii}{2 t}\dd a
=\frac{2}{t}\Re\int_{\ui}^{\uf}\dd u\, m(u),
\end{eqnarray}
where $\ui=\vf+\ii\abar/t$ and $\uf=\vf+\ii\af/t$. Therefore, by $m=\dd p/\dd v$,
\begin{eqnarray}
\label{eq:part2}
F_{\tau=0}(t)&=&
\frac{2}{t}\Re\left[
p\left(\vf+\ii\frac{\af}{t}\right)
-p\left(\vf+\ii\frac{\abar}{t}\right)\right]\qquad (t>0),
\end{eqnarray}
which is a contribution relative to the momentum transfer induced by core-width variation. In going to the second line of Eq.\ (\ref{eq:fjump}) an assumption has been used that $a(\tau)$ varies continuously between values $\ai$ and $\af$ within a time interval shrunk to zero. This amounts to making a constitutive assumption on $a(t)$, indicated by the sign `$\approx$' that stands for a weak equality in the following sense. Consider some function $g(v)$ of $v(t)=\vi\theta(-t)+\vf\theta(t)$. Then $g(v(t))\delta(t)$ has no meaning within Schwartz's theory of distributions, where $\theta(0)$ is undefined. However, a choice
\begin{eqnarray}
g(v(t))\delta(t)\approx\left[\frac{1}{\vf-\vi}\int_{\vi}^{\vf}\dd v\, g(v)\right]\delta(t),
\end{eqnarray}
provides the ``intuitive" result that $\frac{\dd}{\dd t}f(v(t))=f'(v(t))\jump{v}\delta(t)=\delta(t)\int_{\vi}^{\vf}\dd v\,$ $ f'(v)=\jump{f(v)}\delta(t)$. In general, such \emph{weak} definitions of the product $g\delta$ are not unique (Colombeau, 1985) and the adopted one must be warranted by the physical context (Co\-lom\-beau, 1989; 1990). Here, in a pragmatic way, Eq.\ (\ref{eq:part2}) and its consequence, Eq.\ (\ref{eq:selfjump}), will be supported by numerical results in the next Section.

The remaining part of the integral $\int_{-\infty}^t\dd\tau$ is a principal value from which $\tau=0$ is excluded. We split it into parts $\int_{-\infty}^{0^-}\dd\tau$ and $\int_{0^+}^t\dd\tau$, each one being integrated as in Section \ref{sec:counterterm}, but with $a(\tau)\equiv\ai$ and $a(\tau)\equiv\af$, respectively. A simple calculation led as in Section \ref{sec:counterterm} yields the following contribution ($t>0$)):
\begin{eqnarray}
\label{eq:part3}
F^{\pv}(t)&\defi&w_0\frac{2 v(t)}{a(t)\cS}+2\Re\,\pv\int_{-\infty}^t\frac{\dd \tau}{t-\tau}m(\vbar(t,\tau))\frac{\dd \vbar}{\dd\tau}(t,\tau)\\
&=&\frac{2}{\af}\Im\left[ L\left(\vf+\ii\frac{\af}{t}\right)
    -\ii\frac{\af}{t}\frac{L\left(\vf+\ii\frac{\abar}{t}\right)-\!L(\vi\!+\!\ii 0^+)}{\jump{v}+\ii\frac{\abar}{t}}\right]-F_{\tau=0}(t),\nonumber
\end{eqnarray}
where $F_{\tau=0}(t)$ shows up without having to multiply distributions.

Upon gathering Eqs.\ (\ref{eq:part1}), (\ref{eq:part2}) and (\ref{eq:part3}), $F_{\tau=0}$ cancels out so that:
\begin{eqnarray}
F(t)&=&\frac{2}{\ai}\Im L(\vi+\ii 0^+)\quad\text{if}\quad t<0,\nonumber\\
\label{eq:selfjump}
    &=&\frac{2}{\af}\Im\left[ L\left(\vf+\ii\frac{\af}{t}\right)
    -\ii\frac{\af}{t}\frac{L\left(\vf+\ii\frac{\abar}{t}\right)-\!L(\vi\!+\!\ii 0^+)}{\jump{v}+\ii\frac{\abar}{t}}\right]\,\,\text{if}\quad t>0.
\end{eqnarray}
This expression reduces to (\ref{eq:fself}) for $\af=\ai=a$ and $t>0$. The expansion as $\aif/t \to 0$ is identical to Eq.\ (\ref{eq:fselfexpan}) with $a$ replaced by $\af$.
\begin{figure}
\centering
\includegraphics[width=18cm]{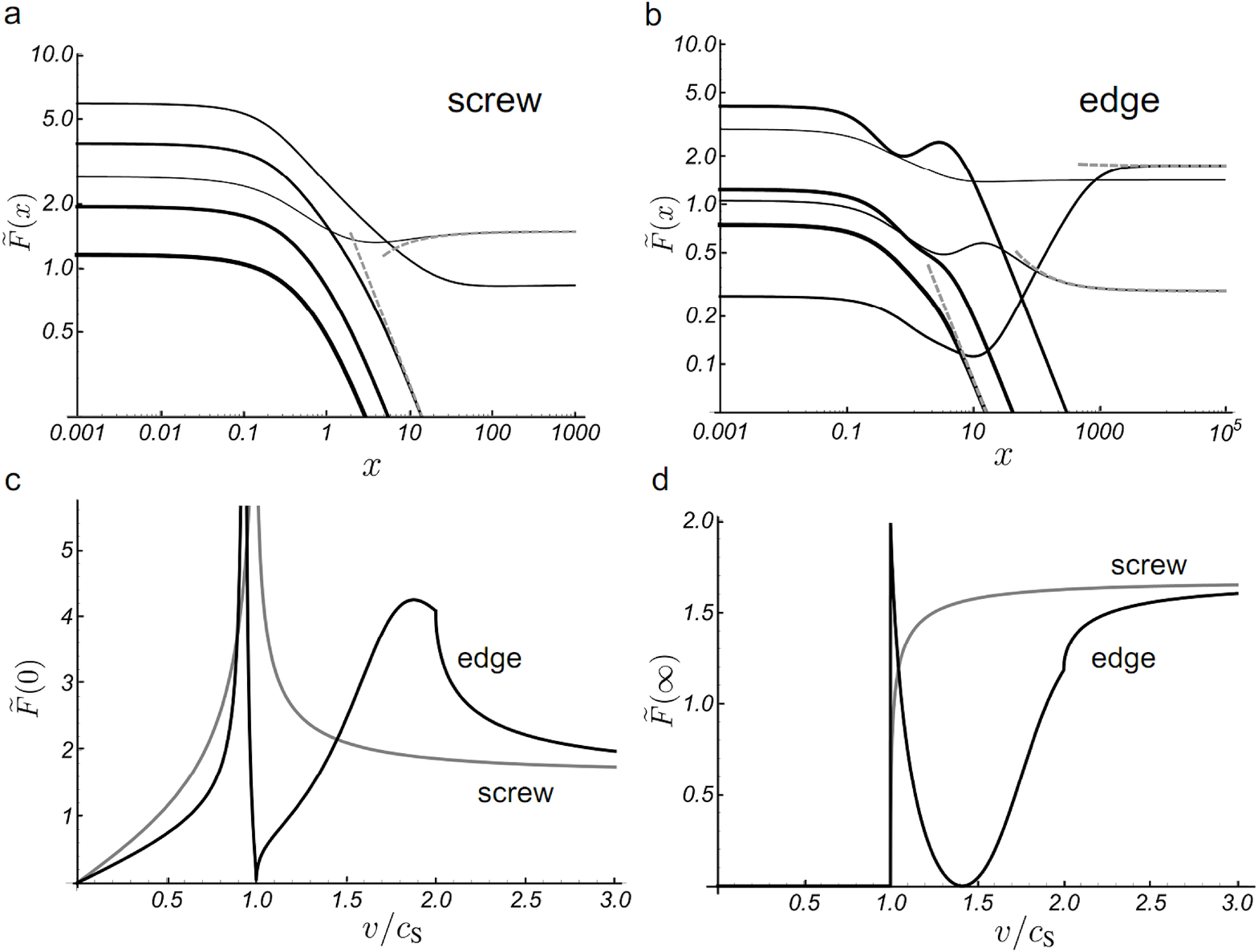}
\caption{\label{fig:fig3} (a) and (b): master curve $\widetilde{F}(x)$ vs.\ $x$ for (a) screw and (b) edge dislocations subjected to a velocity jump from rest to various terminal velocities $\vf$ (see text). Solid: Eq.\ (\ref{eq:selfjump}). The line thickness decreases as $\vf$ increases. Dashed: asymptotic expression (\ref{eq:fselfexpan}). (c) and (d): asymptotic values of $\widetilde{F}(0)$ for zero initial velocity, and $\widetilde{F}(\infty)$ (independent of the initial velocity) as functions of the terminal velocity $v=\vf$. A value $\cL=2\cS$ is used.}
\end{figure}

Since the initial and final widths $\aif$, as computed from $\vif$  by Eq.\ (\ref{eq:av}), are proportional to the interplane distance $d$, the self-force scales as
\begin{eqnarray}
F(t)=w_0 d^{-1} \widetilde{F}(\cS t/d).
\end{eqnarray}
The master curve $\widetilde{F}$, which depends on $\vi/\cS$ and $\vf/\cS$ as parameters, is such that $\widetilde{F}(x)\sim \widetilde{F}(0^+)$ for $x\ll 1$, and $\sim\widetilde{F}(\infty)+\ell x^{-1}$ for $x\gg 1$. The latter regime is described by Eq.\ (\ref{eq:fselfexpan}), from which the analytical expressions of $\widetilde{F}(\infty)$ and the constant $\ell$ are read. One finds for both screw and edge that
\begin{eqnarray}
\label{eq:fzer}
\widetilde{F}(0^+)=\frac{2}{w_0}\frac{d}{\abar}\Im L(\vi+\ii 0^+)+2\frac{d}{\cS}\left(\frac{\vf}{\af}-\frac{\vi}{\abar}\right).
\end{eqnarray}

Figs.\ \ref{fig:fig3}(a) and (b) represent the function $\widetilde{F}(x)$ as a function of $x>0$ in log/log scale, for screws and edges with various terminal velocities. Material parameters $\alpha=0.1$ and $\cL=2\cS$ are used. The initial velocity is $\vi=0$ and the terminal velocities are $\vf=v$, with $v/\cS=0.5$, $0.7$, $0.9$, $1.01$ and $1.2$ for the screw (Fig.\ \ref{fig:fig3}(a)), and the same velocities with $v/\cS=2.1$ added, for the edge (Fig.\ \ref{fig:fig3}(b)). Expression (\ref{eq:av}) of the core width is used. The asymptotic regime given by Eq.\ (\ref{eq:fselfexpan}) is represented for some of the curves as dashed lines. The self-force is seen to strongly depend on $\vf$. A similar conclusion would be drawn upon varying $\vi$, which may lead to negative values of $\widetilde{F}(x)$ (not shown).

Fig.\ \ref{fig:fig3}(b) makes conspicuous a wiggled structure for intermediate values of $x$ in the edge case. The $\bigO(t^{-1})$ term in the asymptotic formula (\ref{eq:fselfexpan}) is seen to be relevant only in a range of high $x$ values in non-subsonic cases. The wild variations of the asymptotic regimes are elucidated in Figs.\ \ref{fig:fig3}(c), (d) where $\widetilde{F}(0^+)$ and $\widetilde{F}(\infty)$ (which is independent of $\vi$) are displayed as functions of the terminal velocity.

Fig.\ \ref{fig:fig3}(c) illustrates the dramatic increase of the self-force at $v=\cS$ (screw) and $v=\cR$ (edge), as the core width shrinks (to a non-zero value due to $\alpha$). At $v=\cS$, the edge core width blows up and the self-force vanishes (see previous section). The quantity $\widetilde{F}(\infty)$ in Fig.\ \ref{fig:fig3}(d) being proportional to the radiative dissipation rate, it vanishes at the radiation-free edge velocity $v_c=\sqrt{2}\cL$ (even for $\alpha>0$ since the phenomenological drag contribution $F_D$ is not part of $F$).

\subsection{Smooth transition in velocity and core width}
\label{sec:logsing}
The $1/d$ scaling of $F(t)$ reported above makes $F(t)$ blow up at the origin in the Volterra limit $d\to 0$. On the other hand and for $|v|<\cS$, $F(t)$ goes to a finite limit for times such that $\cS t/d\gg 1$, owing to the beneficial action of the counter-term examined in Section \ref{sec:counterterm}. This  blowing-up as $d^{-1}$ for $\cS t/d\ll 1$ finds its origin in the infinite acceleration that takes place at $t=0^+$ in the velocity jumps that were considered up to now.

To make some progress, behavior under finite accelerations is studied by spreading over a finite time $\tau_0$ the velocity transition between $\vi$ and $\vf$, taking for instance
\begin{eqnarray}
\label{eq:xispread}
\xi(t)=\vi t+(\vf-\vi)\left[t-\tau_0\left(1-\ee^{-t/\tau_0}\right)\right]\theta(t),
\end{eqnarray}
where accelerated motion begins at position $\xi(0)=0$. The acceleration is now finite, of order $\ddot{\xi}(0)=(\vf-\vi)/\tau_0$. With $v(t)=\dot{\xi}(t)$, we use $a(t)=a(v(t))$
as given by Eq.\ (\ref{eq:av}). The motion reduces to that considered in Section \ref{sec:generalization} as $\tau_0\to 0$. The integral in $F(t)$ over the range $\tau<0$ is done explicitly
as before. One obtains
\begin{subequations}
\label{eq:selfspreadall}
\begin{eqnarray}
F(t)&=&\frac{2}{\ai}\Im L(\vi+\ii 0^+)\qquad\text{for}\qquad t<0,\\
&=&\frac{2w_0}{\cS}\frac{v(t)}{a(t)}+\frac{2}{t}\Re\left[p(v^*(t))
-\frac{L(v^*(t))-L(\vi+\ii 0^+)}{v^*(t)-\vi}\right]\nonumber\\
\label{eq:selfspread}
&&{}+2\Re\int_0^t\frac{\dd\tau}{t-\tau}m(\vbar(t,\tau))\frac{\dd \vbar}{\dd\tau}(t,\tau),
\qquad\text{for}\qquad t>0,\\
v^*(t)&\defi&\frac{\xi(t)+\ii[a(t)+\ai]/2}{t},
\end{eqnarray}
\end{subequations}
where the remaining integral must be evaluated numerically. These equations hold for any $\xi(t)$ at positive times, independently of Eq.\ (\ref{eq:xispread}).

In the following, only velocities $|v|<\cS$ are considered for simplicity, so that  $F(t)=0$ for $t<0$.

A straightforward series expansion shows that the self-force grows up first \emph{linearly} with time, as
\begin{equation}
\label{eq:lingrowth}
F(t)/w_0\simeq \frac{2}{\ai}\left(1-\frac{\vi}{2}\frac{a'(\vi)}{\ai}\right)\frac{\ddot{\xi}(0) t}{\cS}.
\end{equation}
This result holds indifferently for screws and edges, and comes out of the explicit contribution in Eq.\ (\ref{eq:selfspread}), while the integral itself is an $\bigO(t^2)$.

\begin{figure}
\centering
\includegraphics[width=18cm]{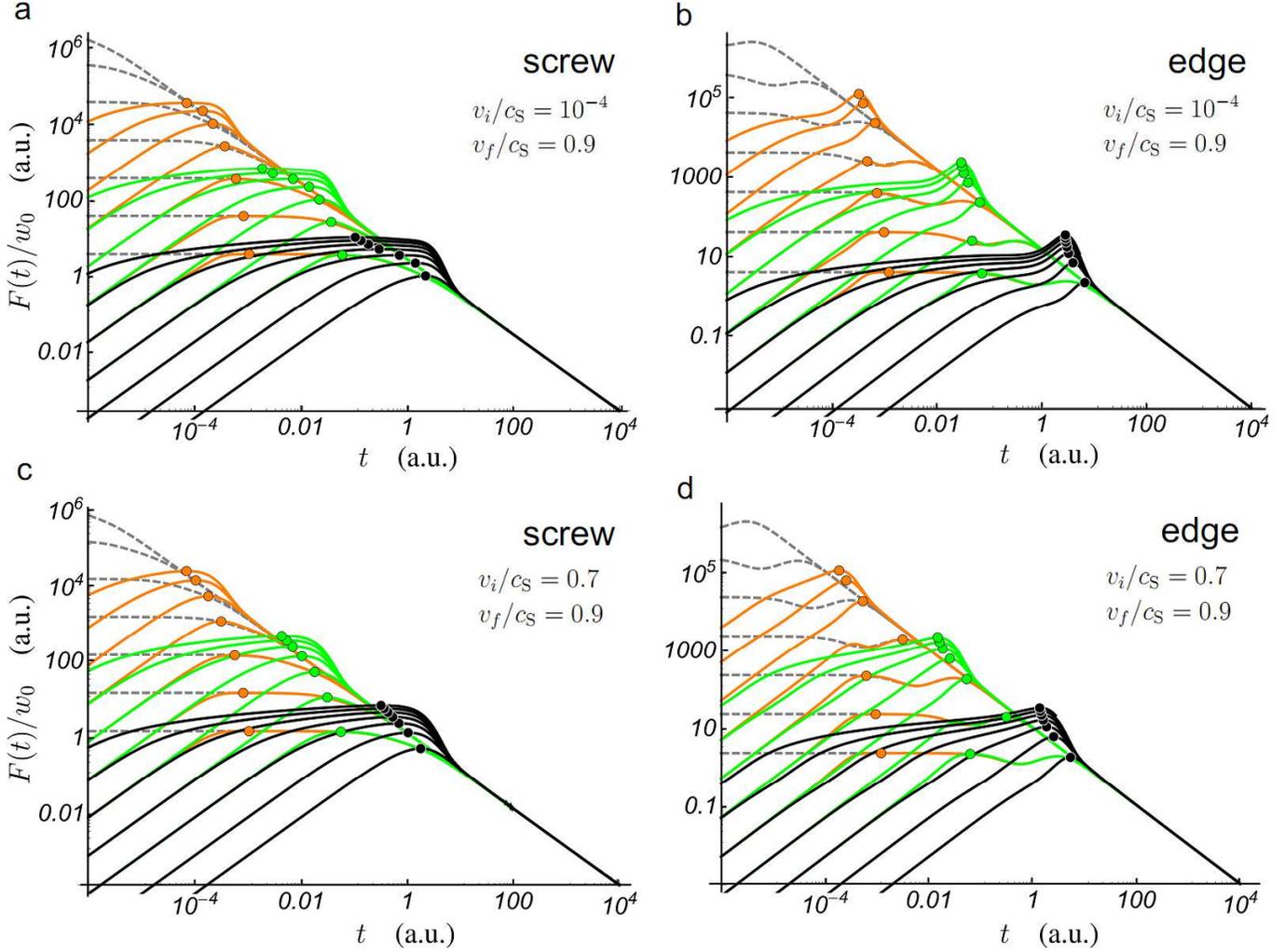}
\caption{\label{fig:fig4} Top figures: self-force $F(t)$ vs.\ time $t$ for (a) screw and (b) edge dislocations subjected to a smooth acceleration over a time $\tau_0$ between velocities $\vi=10^{-4}\cS$ and $\vf=0.9\cS$. The core width varies according to Eq.\ (\ref{eq:av}). The three sets of seven curves correspond to the following values of $\tau_0$ (color online): $5.\ 10^{-3}$ (black, solid), $4.\,10^{-2}$ (green, solid) and $3.\,10^{-1}$ (orange, solid). Within each set (from bottom to top), values of the interplane distance $d=10^{-k}$, $k=1,\ldots,7$ are used. Grey, dashed: asymptotic form (\ref{eq:fselfexpan}) with $a=\af$. Bottom figures: idem, with $\vi$ and $\vf$ as indicated. In all plots $\cL=2\cS$ and $\alpha=0.1$ (arbitrary value). Units are such that $\cS=1$ but otherwise arbitrary (``a.u.'').}
\end{figure}

The dynamical regimes that follow this linear growth step are discussed on the basis of numerical evaluations of $F(t)$ (see Appendix \ref{sec:numcalc}) by means of Eqs.\ (\ref{eq:xispread}) and (\ref{eq:selfspreadall}abc). Figure \ref{fig:fig4} displays $F(t)$ vs.\ $t>0$ for screw and edge dislocations, and two different initial velocities: $\vi=10^{-4}\cS$ in (a) and (b); and $\vi=0.7 \cS$ in (c) and (d). In all plots, $\alpha=0.1$ (arbitrary value), $\cL=2\cS$, and the final velocity is $\vf=0.9\cS$, which lies slightly below the Rayleigh velocity. Four sets of seven curves are represented in log--log scale in each sub-figure. Plots in each set are drawn for exponentially-decreasing values of the interplane distance $d=10^{-k}$, $k=0,\ldots,6$. Each set uses a different value of the characteristic acceleration time $\tau_0$: the grey set (dashed) uses $\tau_0=0$ (infinite acceleration) and represents the jump formulas (\ref{eq:selfjump}); on the other hand, the orange (resp., green, black)\footnote{This order ---grey, orange, green, black--- applies from the top sets to the bottom ones. See online paper for color figures.} set illustrates the effect of a finite acceleration, with $\tau_0=10^{-4}$ (resp., $10^{-2}$, $10^0$). A marker tags the highest value attainted by $F(t)$. The ``edge'' plots may possess up to two local maxima due to the two intervening sound velocities. Also, the dependence in the initial velocity $v_i$ is less conspicuous for the screw than for the edge. In the latter case, $F(t)$ develops a peak for large velocity jumps (Fig.\ \ref{fig:fig4}(b)). Also, while the highest values of $F(t)$ seem to saturate with $1/d$ for the screw, a different behavior takes place for the edge, whose maximal self-force increases more strongly.

\begin{figure}
\centering
\includegraphics[width=18cm]{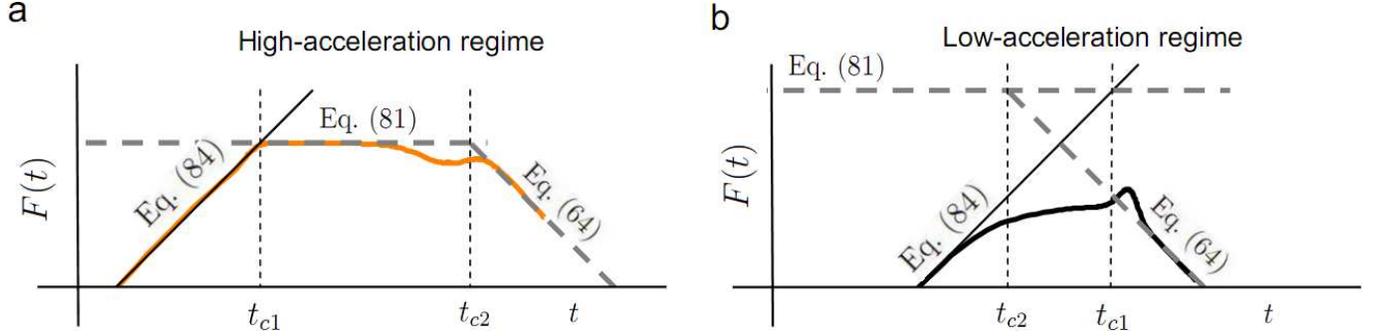}
\caption{\label{fig:fig5} Schematic illustrations of the high- (a) and low-acceleration (b) regimes in log-log scale, as determined by the ordering of crossover times $t_{c1}$ and $t_{c2}$ defined by the intersections of the straight lines. In (a) is reproduced the plot of $F(t)$ vs.\ $t$ of Fig.\ \ref{fig:fig4}(b) with $\tau_0=10^{-4}$ and $d=1$ (orange); in (b), where $t_{c1}$ and $t_{c2}$ have no meaning any longer as crossover times, data of Fig.\ \ref{fig:fig4}(b)  with $\tau_0=1$ and $d=10^{-4}$ (black) are used.}
\end{figure}

In all the solid curves of Fig.\ \ref{fig:fig4}, for which $\tau_0\not =0$, the initial growth regime with slope one is that given by Eq.\ (\ref{eq:lingrowth}). At large times, all curves collapse to the $1/t$ asymptotic regime with slope $-1$ described by Eq.\ (\ref{eq:fselfexpan}) with $\Delta\text{CT}=0$. This asymptote holds for Eq.\ (\ref{eq:selfjump}) as well, see Section \ref{sec:jump}. The intermediate regime between these initial and asymptotic regimes depends on the acceleration $\ddot{\xi}(0)$, as illustrated in Fig.\ \ref{fig:fig5}. At \emph{high accelerations}, i.e., for values of $\tau_0$ small enough, the self-force reaches at intermediate times a plateau, with an undulation in the edge case, of value $\widetilde{F}(0^+)$  given by Eq.\ (\ref{eq:fzer}). This plateau is the short-time regime of the dashed grey curves with $\tau_0=0$, and does not depend on acceleration. Such cases are seen in the bottom curves of the orange and green sets in Fig.\ \ref{fig:fig4}.

The plateau begins at a typical crossover time $t_{c1}$ obtained by matching Eq.\ (\ref{eq:lingrowth}) and (\ref{eq:fzer}), and ends at a crossover time $t_{c2}$ obtained from matching Eq.\ (\ref{eq:fzer}) and the asymptotic Eq.\ (\ref{eq:fselfexpan}), see Fig.\ \ref{fig:fig5}(a). It is easily shown that, for $|\vi|<\cS$,  $t_{c1}$ reduces to $\tau_0$ if $a$ is assumed to be velocity-independent. In the general case, $t_{c1}$ is proportional to $\tau_0$, unlike $t_{c2}$ that depends on $\vif$ (both directly and through $\aif$).

The denomination ``high acceleration'' means that $\tau_0$ is small enough so that $t_{c1}<t_{c2}$. In the opposite case of \emph{low acceleration}, where $t_{c1}>t_{c2}$, the plateau is replaced by a mild increase regime, that may end up with a peak in the edge case, as illustrated by Fig.\ \ref{fig:fig5}(b). Then, $t_{c1,2}$ are not relevant any more as crossover times (see figure). The black curves in Fig.\ \ref{fig:fig4} all belong to the latter case.

The information collected so far for subsonic motion is summarized as follows:
\begin{itemize}
\item For a jump at $t=0$ in both velocity and core width between two steady-state regimes: at times $t\ll d/\cS$, $F$ scales as $1/d$ due to infinite acceleration at jump time. In the limit, the singularity generated at $t=0$ is \emph{not} Dirac-like, but merely that $\propto 1/t$ which arises from the post-jump relaxational asymptote (\ref{eq:cmforce_sub}). As the figures show, this asymptote constitutes an upper bound of $F(t)$ for $\vi=0$. Thus, except \emph{at} jump time, the Volterra limit $d\to 0$ of $F$ is non-singular.

\item For finite accelerations: a ``quasi-jump'' regime is observed, in which the self-force increases in the Volterra limit as for jumps. However, this now takes place within a time range of lower bound $\propto\cS/\ddot\xi(0)$ and of upper bound $\propto d/\cS$, and occurs only for accelerations and $d$ values that are large enough to preserve the order of the bounds. By contrast, for low accelerations or for $d$ small enough, a less-than-$d^{-1}$ increase takes place in the intermediate time regime. Its peculiarity is that peaks show up, \emph{escaping the jump relaxational asymptote}, which is indicative of delayed relaxation.
\end{itemize}

\section{Asymptotic logarithmic behavior}
\label{sec:asym}
Provided that velocities are small, the linearized equations of Section \ref{sec:fdem} are a good approximation, and information about motion at finite times under the action of a constant stress applied at $t=0$, or of an arbitrarily varying stress applied between $t=0$ and some instant $t=t_1$ is extracted (Eshelby, 1953) from Laplace transforms (\ref{eq:laplace}): for both screws and edges, the time-dependent quasimomentum reads $p(t)\simeq m^{\text{eff}}(t)v(t)$, where the function $m^{\text{eff}}(t)$ is $m_0$ times the logarithm in Eqs.\ (\ref{eq:mwslow}) and (\ref{eq:mwelow}) with $|\omega|$ replaced by $1/t$. This is deduced from the small-$s$ expansion of $\wwm(s)$, whose leading term is identical to that in those equations, with $|\omega|$ replaced by $s$. The small-$s$ behavior thus translates into the well-known logarithmic time-dependence of the type $m^{\text{eff}}(t)\propto \log(t/t^*)$, where $t^*$ is a characteristic time proportional to $a$ (i.e., to $d$). The time $t^*$ depends on the dislocation character, but also more generally on the loading conditions (Eshelby, 1953). As far as the linear theory is concerned, it is remarked that since $\tS$ and $\tL$ are proportional to $a$, the small-$s$ expansion of $\wwm$ coincides with its small-$a$ expansion. Thus, Eqs.\ (\ref{eq:flinexpl}) and (\ref{eq:pillon}) are only logarithmically singular in the Volterra limit.

\begin{figure}
\centering
\includegraphics[width=18cm]{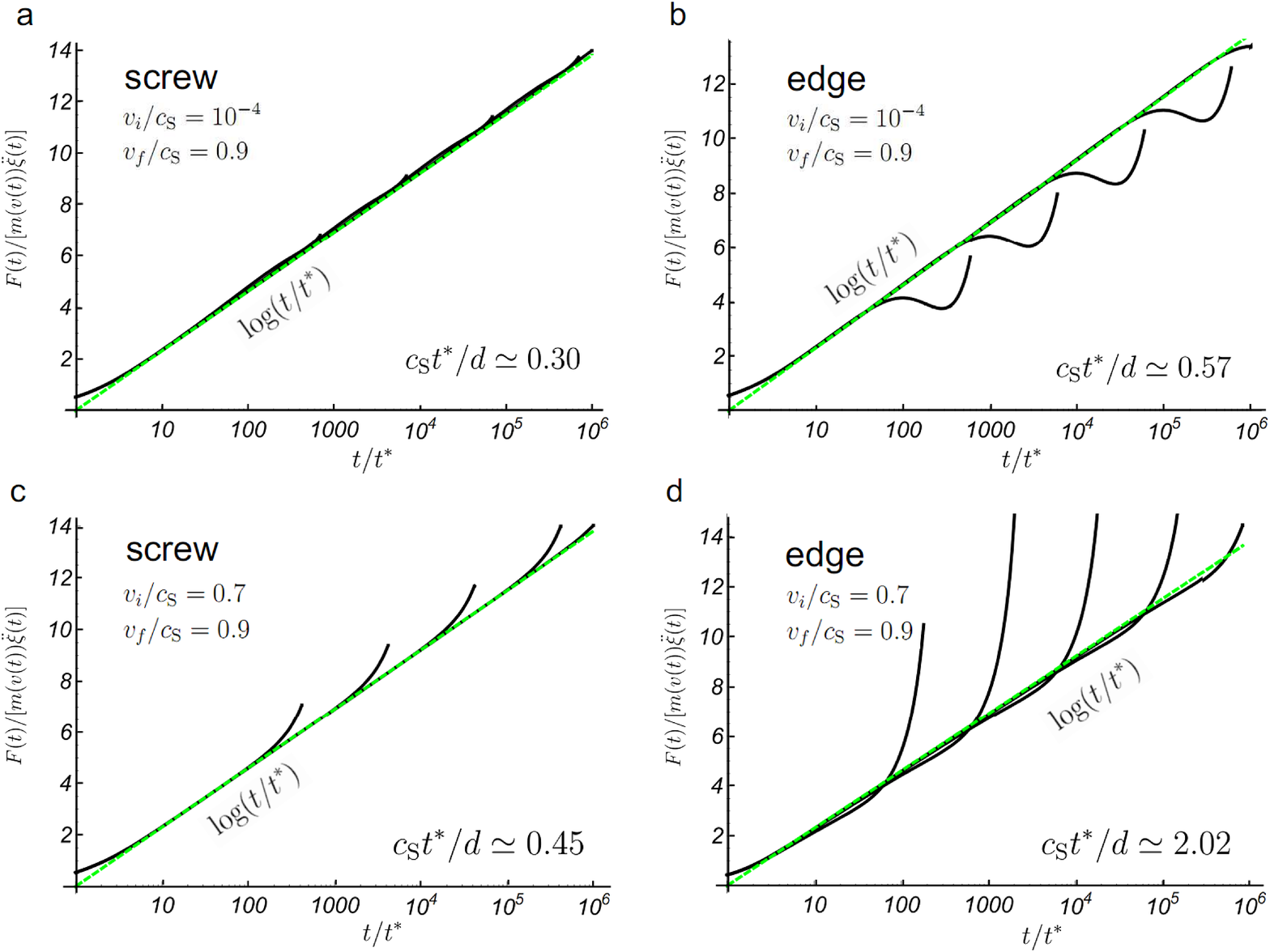}
\caption{\label{fig:fig6} Logarithmic behavior of the self-force in the low-acceleration regime, in log-linear scale. Solid (black): plots for $\cS\tau_0/d=10^2, 10^3,\ldots, 10^6$ (from left to right). Dashed (green): logarithmic asymptotic behavior. Color online.}
\end{figure}
As one allows for arbitrary velocities the linear theory is no more relevant, but extracting the exact asymptotic behavior of the Volterra limit out of the full-fledged self-force (Ni and Markenscoff, 2008) is no easy task. However, one may infer from the latter reference devoted to screw dislocations (and references therein), and from dimensional considerations, that to leading order asymptotic time-behavior of the following type should hold:
\begin{eqnarray}
\label{eq:asympt}
F(t)\simeq m\bigl(v(t)\bigr)\,\log(t/t^*)\,\ddot\xi(t),
\end{eqnarray}
in which $t^*$ is proportional to the core width and \emph{depends on velocity}.

In the subsonic range, the low-acceleration regime is the place to look for, since the previous Section excludes other possibilities. Using the motion of Eq.\ (\ref{eq:xispread}), Fig.\ \ref{fig:fig6} displays in the context of \emph{Model I} log-linear plots of the scaled force $F(t)$ $/[m(v(t))$ $\ddot{\xi}(t)]$ vs.\ $t/t^*$ for $\cS\tau_0/d=10^2,10^3,\ldots 10^6$, which ensures small acceleration conditions. The constant parameters are $\alpha=0.1$ and $\cL/\cS=2$, and the initial and final velocities are as indicated. For better reading, the plots do not extend beyond the crossover time $t_{{\rm c}1}$ introduced in Section \ref{sec:logsing}. The ``overshoots" announce the peaks of Fig.\ \ref{fig:fig4}. Prior to plotting the curves, $t^*$ was determined by plotting the quantity $g(t)\defi t\exp\{-F(t)/[m(v(t))\ddot{\xi}(t)]\}$ for $\cS \tau_0/d=10^6$ against $\cS t/d$ in the range $(1,10^6)$ (not shown). It was found that $g(t)\simeq c_1-c_2 t$, where $c_2$ is a very small coefficient, of order $10^{-6}\sim 10^{-7}$. While this confirms that $F(t)$ behaves as  $\log(\cS t/d)$, it is not clear to this author whether the observed systematic variation of $t^*$ with time is a genuine effect or a numerical artifact due to precision loss. In practice, $t^*$ was identified to $c_1$, which is sufficient to ensure good overall scaling.

The plots indicate that $t^*$ depend on velocity. Moreover, it is observed that even for very small initial velocities, the determined values of $\cS t^*/d$, indicated in the plots, markedly differ from those provided by the leading-order logarithm of the Fourier (or Laplace) mass response function. The latter is read from Eqs.\ (\ref{eq:mwslow}) and (\ref{eq:mwe2}). Using the appropriate values of $a=a(0)$ given after Eqs.\ (\ref{eq:av}b), this gives $\cS t^*/d\simeq 0.540$ (screw) and $1.012$ (edge), respectively. To exclude the possibility of an error in the code, $F(t)$ was computed for $\vi=0$, $\vf=10^{-4}$ both by direct numerical integration over time using the full integral formula (see Appendix \ref{sec:numcalc}), and by numerical Laplace inversion of the linearized equation $F^{\text{lin}}(s)=s^2\wwm(s)\xi(s)$ by means of a contour that runs on both sides of the negative-axis cut, skirting round the origin. Same results of $F(t)$ were found to four significant digits for screw and edges, which validates both procedures. It ensues that the characteristic time provided by the FT or LT is irrelevant to general applications, and that a theoretical expression (or other means of computation) of $t^*$ that reproduces the values indicated in the plots should be looked for. For screws, the expression provided in Ni and Markenscoff (2008) is a possible candidate, but a detailed study of this issue lies outside the scope of the present work.

\section{Conclusion}
\label{sec:concl}
Starting from exact dynamical core equations for screw and edge dislocations in an isotropic medium, equations of motion (EoM) were deduced by means of an arctangent \textit{ansatz} with arbitrary time variation of the core width. The stationary line energy density function $W(v)$ was shown to uniquely determine the self-force, with same structure in both cases, Eq.\ (\ref{eq:fesh_screw}).
They contain a special local term the regularizing nature of
which was made conspicuous. This term naturally arises from the formalism
employed (Pellegrini, 2010). It is distinctive of the present theory,
although its concealed presence can be traced in the EoM for screw
dislocations derived by Eshelby (1953) from an electromagnetic analogy.
For $|v|<\cS$ in the volterra limit, this term was shown  to cancel out an otherwise diverging contribution in the self-force. An expression for it in terms of the energy, Eq.\ (\ref{eq:ctv2}), was proposed.

The arctan \textit{ansatz} evidently simplifies matters, allowing one to introduce a complex ``velocity", of real part an average velocity, and of imaginary part an average core width divided by the averaging time interval, which is perhaps not too surprising in view of the analytic structure of static PN solutions (Lej\v{c}ek, 1976). This device, which consistently handles the square-root branch cut, makes the obtained self-force expressions formally hold for all velocity regimes, as has been shown for steady motion in Section \ref{sec:tdcw}.

Specific analytical results were obtained: the self-force $F(t)$ associated to uniformly moving
screw and edge dislocations that undergo an instantaneous velocity change,
Eq.\ (\ref{eq:fself}); its generalization to instantaneous changes in core width, Eq.\ (\ref{eq:selfjump}); the fre\-quen\-cy-dep\-end\-ent mass and damping function of the edge dislocation, Eqs.\ (\ref{eq:mwe}) and (\ref{eq:alpe}); Eq.\ (\ref{eq:defABC}), which provides a simple definition of the important functions $A(v)$ and $B(v)$ that enter Weertman's equations (1969); the Rosakis (2001) steady-state equations in terms of the Lagrangian. Although some of these features are relevant to velocities $|v|>\cS$, applications to this regime should be regarded with caution for reasons evoked in Section \ref{sec:tdcw}. Finally, a synthetic expression of $F(t)$ for arbitrary motion from a state of constant velocity was given in Eq.\ (\ref{eq:selfspreadall}).

To support these findings, various checks the EoM and of the underlying formalism were performed, allowing one to retrieve a number of known results as particular cases: in Section \ref{sec:lossterm}, results by Markenscoff (1980) and Callias and Markenscoff (1980) concerning the self-stress on the slip plane produced by a Volterra dislocations; as Eq.\ (\ref{eq:cmforce_sub}), the self-force for subsonic smeared-out screw and edge dislocations jumping from rest to constant velocity, in the limit of vanishing core width (Clifton and Markenscoff, 1981); in Section \ref{sec:linkesh},  Eshelby's EoM for screws (1953); in Section \ref{sec:fdem}, the frequency-dependent mass of the screw dislocation in oscillatory motion at small velocities (Pillon et al., 2007); in Section \ref{sec:tdcw} the steady-state EoM of \emph{Model I} (Rosakis, 2001). All of these stand as indirect verifications of the dynamical core equations reviewed in Section \ref{sec:DynPN}.

Finally, a distinction was introduced in the subsonic range between high- and low-acceleration regimes on the basis of a quantitative criterion, $\log t$ behavior of the self-force being observable for low accelerations only, associated to a non-trivial characteristic time about which we gave some information of numerical nature. No similar analysis was undertaken in the context of sonic transitions, for lack of a proper dynamical core-width model.

Our generic expressions of $F(t)$, which at least apply to screw and edge dislocations in an isotropic medium, might extend to anisotropic media as well (Bullough and Bilby, {\color{red} 1954}; Teutonico, 1961), although this remains in need of a formal demonstration. Obviously, the underlying structure revealed by our analysis (e.g., the occurrence of a complex Lagrangian in connection with dissipation in Mach fronts) calls for deeper understanding.

\section{Acknowledgements}
The author thanks C.\ Denoual for enjoyable discussions in the preliminary steps, and a reviewer for comments. A discussion with X.\ Markenscoff prompted the writing of Section \ref{sec:lossterm}.
\appendix
\numberwithin{equation}{section}
\section{Energies and related matters}
\label{eq:remid}
The velocity-dependent kinetic energy, $\Wk$, strain energy, $\Ws$, and total energy $W=\Wk+\Ws$, of the displacement field associated to uniformly moving screw and edge dislocations have been computed by Weertman (1961), after early steps were taken by Frank (1949) and Eshelby (1949). See Weertman and Weertman (1980) and Lothe (1992) for reviews.

Let $\bSL\defi\sqrt{1-(v/\cSL)^2}$. In notations close to that of  Hirth et al.\ (1998) one has, for the screw dislocation,
\begin{subequations}
\label{eq:wsall}
\begin{eqnarray}
\label{eq:wks}
\Wk(v)&=&\frac{w_0}{2}\left(-\bS+\bS^{-1}\right),\\
\label{eq:wss}
\Ws(v)&=&\frac{w_0}{2}\left(\bS+\bS^{-1}\right),\\
\label{eq:ws}
W(v)&=&w_0\bS^{-1},\text{ with } W(0)=w_0,
\end{eqnarray}
\end{subequations}
and for the edge dislocation
\begin{subequations}
\label{eq:weall}
\begin{eqnarray}
\label{eq:wke}
\Wk(v)&=&\frac{w_0}{2}\left(\frac{\cS}{v}\right)^2\left(4\bL+4\bL^{-1}+\bS^3
-5\bS-5\bS^{-1}+\bS^{-3}\right),\\
\label{eq:wse}
\Ws(v)&=&\frac{w_0}{2}\left(\frac{\cS}{v}\right)^2\left(12\bL+4\bL^{-1}-\bS^3
-9\bS-7\bS^{-1}+\bS^{-3}\right),\\
\label{eq:we}
W(v)&=&\frac{w_0}{2}\left(\frac{\cS}{v}\right)^2\left(16\bL+8\bL^{-1}-14\bS
-12\bS^{-1}+2\bS^{-3}\right)\\
\label{eq:wo}
&&\text{ with } W(0)=\frac{w_0}{1-\nu}=2w_0\left(1-\frac{\cS^2}{\cL^2}\right),
\end{eqnarray}
\end{subequations}
where $\nu$ is Poisson's ratio. In statics, the quantity $w_0\defi\log(R/r_0)$ $(\mu b^2)$ $/(4\pi)$ is the rest energy factor. The logarithm $\log(R/r_0)$, where $R$ and $r_0$ are the outer and inner cut-off radii, does not enter the dynamic problem (Beltz.\ et al., 1968), and the identities to be given are independent of it. Accordingly, we use everywhere $w_0$ as defined by Eq.\ (\ref{eq:w0defmain}); whenever the terms ``energies'' or ``Lagrangian'' are employed, they refer to prelogarithmic terms only.

In the text, we need the limit of $\Wk(v)$ as $v\to\infty$ with $\Im v>0$, and the order of the next term. Because with the principal determination of the square root,
$\sqrt{-z^2}=-\ii z\sign\Im z$ for
$z\in\mathbb{C}\setminus\mathbb{R}$, one has in this limit $\bSL\sim -\ii (v/\cSL)$.
Owing to the term $-\bS$ for the screw, and to the term $\bS^3/v^2$ for the edge, the above expressions yield for both
\begin{eqnarray}
\label{eq:limit}
\Wk(v)=\ii\frac{w_0}{2}\frac{v}{\cS}+\bigO(1/v),
\end{eqnarray}

The ``stationary Lagrangian'' is (Stroh, 1962; Beltz et al., 1968; Hirth, Zbib and Lothe, 1998)
\begin{eqnarray}
\label{eq:lagdef}
L(v)\defi \Wk(v)-\Ws(v)=2\Wk(v)-W(v).
\end{eqnarray}
Introducing ${\bS}_2\defi\sqrt{1-v^2/(2\cS^2)}$, Eqs.\ (\ref{eq:wsall}ab) and (\ref{eq:weall}ab) provide (Beltz et al., 1968)
\begin{subequations}
\label{eq:lagse}
\begin{eqnarray}
L(v)&=&-w_0 \bS\hspace{4.6cm} (\text{screw}),\\
    &=&-4w_0\left(\frac{\cS}{v}\right)^2\left(\bL-\bS^{-1}{\bS}_2^4\right)\qquad (\text{edge}),
\end{eqnarray}
\end{subequations}
so that $L(0)=-w_0$ (screw) and $L(0)=-w_0/(1-\nu)$ (edge). The edge Lagrangian vanishes at the Rayleigh velocity where the kinetic and potential energies coincide (Teutonico, 1961).

A quasimomentum is introduced as $p=\dd L/\dd v$ (e.g., Hirth et al., 1998). The following identity was demonstrated by Beltz et al.\ (1968) from the volume-integral expression of $L(v)$:
\begin{eqnarray}
\label{eq:beltz}
W_k=\frac{1}{2}p\,v.
\end{eqnarray}
Thus $p(v)$ is readily deduced from the explicit expressions of $W_k$, and matches Eqs.\ (\ref{eq:fsvdef}) and  (\ref{eq:fevdef}). Introducing a mass function as $m\defi\dd p/\dd v$ (other definitions are found in the literature, e.g. Weertman, 1961; Sakamoto, 1991; Ni and Markenscoff, 2008), it follows that
\begin{eqnarray}
\label{eq:massident1}
m=\frac{\dd p}{\dd v}=\frac{\dd^2 L}{\dd v^2}=2\frac{\dd}{\dd v}\left(\frac{W_k}{v}\right).
\end{eqnarray}
From the above, one also sees that (e.g., Hirth et al., 1998)
\begin{eqnarray}
\label{eq:massident2}
m=\frac{1}{v}\frac{\dd}{\dd v}(pv-L)=\frac{1}{v}\frac{\dd W}{\dd v},
\end{eqnarray}
where the first equality makes use of the definitions $p=\dd L/\dd v$ and $m=\dd p/\dd v$, and the second one appeals to Eqs.\  (\ref{eq:lagdef}) and (\ref{eq:beltz}). The right-hand side complies with Frank's (1949) definition of $p$, such that
$\dd p/\dd t=\dd W/\dd x$ ($x$ being the position), which implies in the steady state that
\begin{eqnarray}
\label{eq:pxv}
p=\int_0^v\frac{\dd v}{v}\frac{\dd W}{\dd v}.
\end{eqnarray}
Gathering Eqs.\ (\ref{eq:massident1}) and (\ref{eq:massident2}) leads to the differential relationship
\begin{eqnarray}
\label{eq:identwkw}
2\frac{\dd}{\dd v}\left(\frac{W_k}{v}\right)=\frac{1}{v}\frac{\dd W}{\dd v}.
\end{eqnarray}
Finally, as in relativistic mechanics of mass points (e.g., Bergmann, 1976, p.\ 92), the kinetic energy of the dislocation is the velocity-dependent part of its energy
\begin{eqnarray}
\label{eq:Kvm}
K(v)\defi W(v)-W(0)=\int_0^v \dd v\,\frac{\dd W}{\dd v}(v)=\int_0^v \dd v\,m(v) v.
\end{eqnarray}
\section{Useful integrals}
\label{sec:useint}
Some integrals used in the calculations of Section \ref{sec:selfforce} are listed hereafter, where it is assumed that $\Im(v)>0$, with $\Re(v)$ arbitrary. Principal determinations of the functions are used as defined in Abramowitz and Stegun (1972).
\begin{subequations}
\begin{eqnarray}
\label{eq:b1}
&&\int_{-1}^1 \frac{\dd u}{\pi} \frac{u}{\sqrt{1-u^2}}\frac{1}{u-v}=1+i\frac{v}{\sqrt{1-v^2}},\\
\label{eq:b2}
&&\int_{-1}^1 \frac{\dd u}{\pi} \frac{1}{\sqrt{1-u^2}}\frac{1}{(u-v)^2}=i\frac{v}{(1-v^2)^{3/2}},\\
&&\int_{-1}^1 \frac{\dd u}{\pi}\frac{1}{u^3} \left(\frac{2-u^2}{\sqrt{1-u^2}}-2\right)\frac{1}{u-v}=\frac{1}{v^3}\left[
i\frac{2-v^2}{\sqrt{1-v^2}}-\frac{4}{\pi}\left(v-\text{coth}^{-1}v\right)\right],\nonumber\\
\label{eq:b3}
&&\\
&&\int_{\cS<|u|<\cL} \frac{\dd u}{u^3}\frac{1}{u-v}=\frac{2}{v^3}
\left[\left(\frac{v}{\cL}-\text{tanh}^{-1}\frac{v}{\cL}\right)-\left(\frac{v}{\cS}
-\text{tanh}^{-1}\frac{v}{\cS}\right)\right],\nonumber\\
&&\hspace{10cm} (0<\cS<\cL),\nonumber\\
\label{eq:b4}
&&\hspace{3.25cm}{}=\frac{2}{v^3}
\left[\left(\frac{v}{\cL}-\text{coth}^{-1}\frac{v}{\cL}\right)-\left(\frac{v}{\cS}
-\text{coth}^{-1}\frac{v}{\cS}\right)\right].
\end{eqnarray}
\end{subequations}
It is recalled (see above reference) that $\text{tanh}^{-1}z=\text{coth}^{-1}z+i(\pi/2)\sign(\Im z)$.

The following series (Gradshteyn and Ryzhik, 2007) serves to evaluate the Fourier transforms over time in Section \ref{sec:fdem}. We use the notation $I\!\mathbf{L}_\nu(z)\defi {\frac{\pi}{2}}[I_\nu(z)-\mathbf{L}_\nu(z)]$, and $K_\nu$, $I_\nu$ and $\mathbf{L}_\nu$ are the modified Bessel and Struve functions (e.g., Abramowitz and Stegun, 1972).
\begin{subequations}
\begin{eqnarray}
\label{eq:c1}
&&\int_0^\infty \frac{\ee^{\ii\omega t}\dd t }{(1+t^2)^{5/2}}=\frac{\omega^2}{3}K_2(|\omega|)+\ii\frac{\omega}{3}\left[1-\frac{\omega^2}{3}+|\omega|\,I\!\mathbf{L}_2(|\omega|)\right],\\
\label{eq:c2}
&&\int_0^\infty \frac{\ee^{\ii\omega t}\dd t }{(1+t^2)^{3/2}}=|\omega|K_1(|\omega|)+\ii \omega\bigl[1-I\!\mathbf{L}_1(|\omega|)\bigr],\\
\label{eq:c3}
&&\int_0^\infty \frac{\ee^{\ii\omega t}\dd t }{(1+t^2)^{1/2}}=K_0(|\omega|)+\ii\sign(\omega)I\!\mathbf{L}_0(|\omega|),\\
\label{eq:c4}
&&\int_0^\infty \left(\sqrt{1+t^2}-t\right)\ee^{\ii\omega t}\dd t=\frac{1}{\omega^2}-\frac{K_1(|\omega|)}{|\omega|}+\frac{\ii}{\omega}I\!\mathbf{L}_1(|\omega|).
\end{eqnarray}
\end{subequations}
As functions of $v$ in the upper complex plane, integrals (\ref{eq:b1})--(\ref{eq:b4}) have been checked numerically with the \emph{Mathematica} software (Wolfram Research, 2007). A similar check was carried out for integrals  (\ref{eq:c1})--(\ref{eq:c4}) as functions of $\omega\in\mathbb{R}$.

\section{Numerical method}
\label{sec:numcalc}
The integral in Eq.\ (\ref{eq:selfspread}) is evaluated with \emph{Mathematica}'s general-purpose adaptative integration routine (Wolfram Research, 2007), and a change of variable $\tau=t-\exp(-u)$ with $u$ running from $-\log t$ to $+\infty$. Beforehand, Eqs.\ (\ref{eq:selfspreadall}) were compared numerically with the more general Eq.\ (\ref{eq:fesh_screw}) for various test cases including negative times, and decelerated motion from supersonic to subsonic velocities. Equation (\ref{eq:fesh_screw}) was integrated with $u$ from $-\infty$ to $+\infty$ with $u=-\log t$ specified as a possible numerically problematic point when $t>0$. The reduction to a finite integration range was left to \emph{Mathematica}. Same numerical values were found, to seven significant digits at least in all cases examined. Prescriptions such as $v+\ii 0^+$ in Eqs.\ (\ref{eq:asig}ab) are implemented with $0^+\lesssim 10^{-9}$.

\end{document}